\pdfoutput=1
\documentclass[12pt,a4paper]{article}
\usepackage{ifthen} 
\newboolean{pdflatex}
\setboolean{pdflatex}{true} 

\newboolean{articletitles}
\setboolean{articletitles}{false} 

\newboolean{uprightparticles}
\setboolean{uprightparticles}{false} 

\newboolean{inbibliography}
\setboolean{inbibliography}{true} 

\usepackage{microtype}
\usepackage{lineno}  
\usepackage{xspace} 
\usepackage{caption} 
\usepackage[utf8]{inputenc}

\usepackage{graphicx}  
\usepackage{color}
\usepackage{colortbl}
\graphicspath{{./figs/}} 
\usepackage{xcolor}
\usepackage[most]{tcolorbox}
\usepackage[percent]{overpic}
\usepackage[varg]{txfonts}
\usepackage{relsize}

\usepackage{amsmath} 
\usepackage{amssymb}
\usepackage{amsfonts}
\usepackage{upgreek} 
\usepackage{cite}
\DeclareUnicodeCharacter{2217}{ }

\newcommand*\patchAmsMathEnvironmentForLineno[1]{%
\expandafter\let\csname old#1\expandafter\endcsname\csname #1\endcsname
\expandafter\let\csname oldend#1\expandafter\endcsname\csname
end#1\endcsname
 \renewenvironment{#1}%
   {\linenomath\csname old#1\endcsname}%
   {\csname oldend#1\endcsname\endlinenomath}%
}
\newcommand*\patchBothAmsMathEnvironmentsForLineno[1]{%
  \patchAmsMathEnvironmentForLineno{#1}%
  \patchAmsMathEnvironmentForLineno{#1*}%
}
\AtBeginDocument{%
\patchBothAmsMathEnvironmentsForLineno{equation}%
\patchBothAmsMathEnvironmentsForLineno{align}%
\patchBothAmsMathEnvironmentsForLineno{flalign}%
\patchBothAmsMathEnvironmentsForLineno{alignat}%
\patchBothAmsMathEnvironmentsForLineno{gather}%
\patchBothAmsMathEnvironmentsForLineno{multline}%
\patchBothAmsMathEnvironmentsForLineno{eqnarray}%
}

\usepackage{color} 
\graphicspath{{./}} 

\def\Y#1S{\ensuremath{\Upsilon{(#1S)}}\xspace}
\def\OneS  {\Y1S}
\def\TwoS  {\Y2S}
\def\ThreeS{\Y3S}
\def\piz   {\ensuremath{\pi^0}\xspace}
\def\pip   {\ensuremath{\pi^+}\xspace}
\def\pim   {\ensuremath{\pi^-}\xspace}
\def\Kp    {\ensuremath{K^+}\xspace}
\def\Km    {\ensuremath{K^-}\xspace}
\newcommand{\al}{\ensuremath{\kern 0.5em }}
\newcommand{\all}{\ensuremath{\kern 0.25em }}
\newcommand{\gkk}{\ensuremath{\gamma\Kp\Km}\xspace}
\newcommand{\gpipi}{\ensuremath{\gamma\pip\pim}\xspace}
\def\babar{BaBar}
\def\epem       {\ensuremath{e^+e^-}\xspace}
\newcommand{\gev}{\ensuremath{\mathrm{\,Ge\kern -0.1em V}}\xspace}
\newcommand{\mev}{\ensuremath{\mathrm{\,Me\kern -0.1em V}}\xspace}
\newcommand{\mevcc}{\ensuremath{{\mathrm{\,Me\kern -0.1em V\!/}c^2}}\xspace}
\newcommand{\gevcc}{\ensuremath{{\mathrm{\,Ge\kern -0.1em V\!/}c^2}}\xspace}
\renewcommand{\gg}{\ensuremath{\gamma\gamma}}
\newcommand{\etaprkk}{\ensuremath{\etapr \Kp \Km}\xspace}
\newcommand{\etaprpipi}{\ensuremath{\etapr \pip \pim}\xspace}
\newcommand{\etapipi}{\ensuremath{\eta \pip \pim}\xspace}

\newcommand{\etac}{\ensuremath{\eta_{c}}\xspace}
\newcommand{\etapr}{\ensuremath{\eta^{\prime}}\xspace}
\def\invfb   {\ensuremath{\mbox{\,fb}^{-1}}\xspace}
\def\jpsi     {\ensuremath{{J\mskip -3mu/\mskip -2mu\psi\mskip 2mu}}\xspace}
\def\etal{{\em et al.}}
\def\BR         {{\ensuremath{\cal B}\xspace}}
\def\calB         {{\ensuremath{\cal B}\xspace}}

\def\calR         {{\ensuremath{\cal R}\xspace}}

\def\calL         {{\ensuremath{\cal L}\xspace}}

\newcommand{\pips}{\ensuremath{\pi_s^+}\xspace}
\newcommand{\pims}{\ensuremath{\pi_s^-}\xspace}
\newcommand{\mm}{\ensuremath{M_{\mathrm{rec}}}\xspace}
\newcommand{\mmtwo}{\ensuremath{M^2_{\mathrm{rec}}}\xspace}
\def\ep         {\ensuremath{e^+}\xspace}
\def\en         {\ensuremath{e^-}\xspace}
\def\mup        {\ensuremath{\mu^+}\xspace}
\def\mun        {\ensuremath{\mu^-}\xspace}
\def\pt         {\mbox{$p_T$}\xspace}
\def\gevc{\mbox{${\mathrm{GeV}}/c\ $}}
\def\KS    {\ensuremath{K^0_{\scriptscriptstyle S}}\xspace}
\def\gevccc{\mbox{${\mathrm{GeV^2}}/c^4$}}
\newcommand{\almm}{\ensuremath{\kern -1.00em }}
\newcommand{\aln}{\ensuremath{\kern -0.25em }}
\newcommand{\alm}{\ensuremath{\kern -0.50em }}
\def\pipm  {\ensuremath{\pi^\pm}\xspace}

\begin{document}

\begin{titlepage}

\pagenumbering{roman}

\vspace*{2.0cm}

{\normalfont\bfseries\boldmath\large
\begin{center}
Light meson spectroscopy and gluonium searches in $\eta_c$ and $\Upsilon(1S)$ decays at \babar
\end{center}
}

\vspace*{1.0cm}

\begin{center}
  Antimo Palano, \\
  INFN Sezione di Bari, Italy\\
\end{center}
\begin{abstract}
  \noindent
  We study the \OneS radiative decays to \gpipi and \gkk using data 
recorded with the \babar\ detector operating at the SLAC PEP-II
asymmetric-energy \epem\ collider at center-of-mass energies at the
\TwoS and \ThreeS resonances.
The \OneS resonance is reconstructed from the decay 
$\Upsilon(nS)\to \pi^+ \pi^- \OneS$, $n=2,3$.
We also study the processes $\gg\to\etac\to\etaprkk$, $\etaprpipi$, and $\etapipi$ using a data sample
of 519~\invfb\ recorded with the \babar\ detector at center-of-mass energies at and near the
$\Upsilon(nS)$ ($n = 2,3,4$) resonances.
A Dalitz plot analysis is performed of \etac decays to $\etapr \Kp \Km$, $\etapr \pip \pim$, and $\eta \pip \pim$.
A new $a_0(1700)$ resonance is observed in the $\eta \pipm$ invariant-mass spectrum from the $\etac\to \etapipi$ decay.
We compare \etac decays to $\eta$ and \etapr final states in association with scalar mesons as they relate to the identification of the scalar glueball.
\end{abstract}
\vspace*{2.0cm}

{\normalfont\bfseries\boldmath
\begin{center}
  On behalf of the \babar\ Collaboration\\
  Presented at the ``Eighth Workshop on Theory, Phenomenology and Experiments in Flavour Physics''-FPCapri2022\\
  {\it Jun 11 – 13, 2022, Villa Orlandi, Anacapri, Capri Island, Italy}
\end{center}
}

\end{titlepage}

\section{Introduction}
The existence of gluonium states is still an open issue for
Quantum Chromodynamics (QCD). Lattice QCD calculations predict the lightest gluonium states
to have quantum numbers $J^{PC}=0^{++}$ and $2^{++}$ and to be in
the mass region below 2.5 \gevcc~\cite{Chen:2005mg}.
In particular, the $J^{PC}=0^{++}$ glueball is predicted to have a mass around 1.7 \gevcc.
The broad $f_0(500)$, $f_0(1370)$~\cite{Minkowski:1998mf}, $f_0(1500)$~\cite{Amsler:1995tu,Amsler:1995td}, 
$f_0(1710)$~\cite{Janowski:2014ppa,Gui:2012gx} and possibly the $f_0(2100)$~\cite{Ablikim:2013hq} have been suggested as scalar glueball candidates.
However, the identification of the scalar glueball is complicated by the possible mixing with standard $q \bar q$ states.

Radiative decays of heavy quarkonia, in which a photon replaces one of the three gluons
from the strong decay of \jpsi or \OneS, can probe color-singlet
two-gluon systems that produce gluonic resonances.
\jpsi decays
have been extensively studied~\cite{Kopke:1988cs,Dobbs:2015dwa}.
In the first \babar\ analysis~\cite{BaBar:2018uqa} summarized in the present review, we study \OneS decays,
taking into account that
the experimental observation of radiative \OneS decays is challenging because their rate
is suppressed by a factor of $\approx 0.025$ compared
to \jpsi radiative decays, which are of order $10^{-3}$~\cite{CLEO:2005koa}.%

Decays of the \etac, the lightest pseudoscalar $c \bar{c}$ state, provide a window on light meson states.
In the second analysis~\cite{BaBar:2021fkz} summarized in the present review, we consider the three-body \etac decays to \etaprkk, \etaprpipi, and \etapipi ,  
using two-photon interactions, $\epem\to\epem\gamma^*\gamma^*\to\epem\etac$.
If both of the virtual photons are quasi-real,
the allowed $J^{PC}$ values of any produced resonances are $0^{\pm+}$, $2^{\pm+}$, $4^{\pm+}$...~\cite{Yang:1950rg}.       
The possible presence of a gluonic component of the \etapr meson, due to the so-called gluon anomaly, has been discussed in recent years~\cite{Harland-Lang:2013ncy,Bass:2018xmz}. 
A comparison of the $\eta$ and \etapr content of \etac decays might yield information on the possible gluonic content of resonances decaying to $\pip \pim$ or $\Kp \Km$.

\section{Study of \OneS radiative decays to $\gamma \pip \pim$ and $\gamma \Kp \Km$}

\subsection{Events reconstruction}
\label{sec1}
We reconstruct the decay chains
\begin{equation}
  \TwoS/\ThreeS \to (\pips \pims)  \OneS \to (\pips \pims) (\gamma \pip \pim) 
\label{eq:twoa}
\end{equation}
and
\begin{equation}
  \TwoS/\ThreeS \to (\pips \pims)  \OneS  \to (\pips \pims) (\gamma \Kp \Km),
\label{eq:twob}
\end{equation}
where we label with the subscript $s$ the slow pions from the
direct \TwoS and \ThreeS decays.

\begin{figure}
  \centering
  \includegraphics[width=7cm]{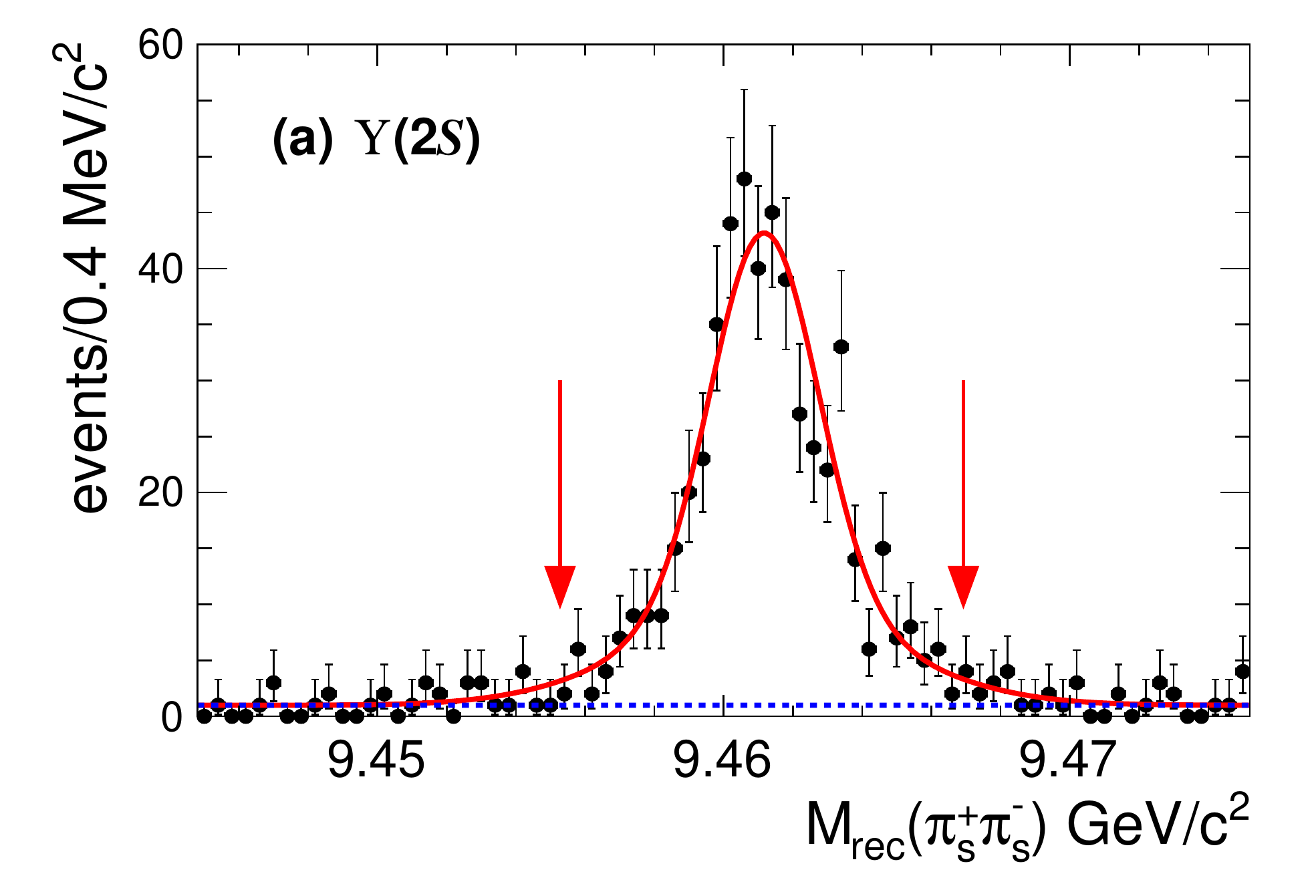}
  \includegraphics[width=7cm]{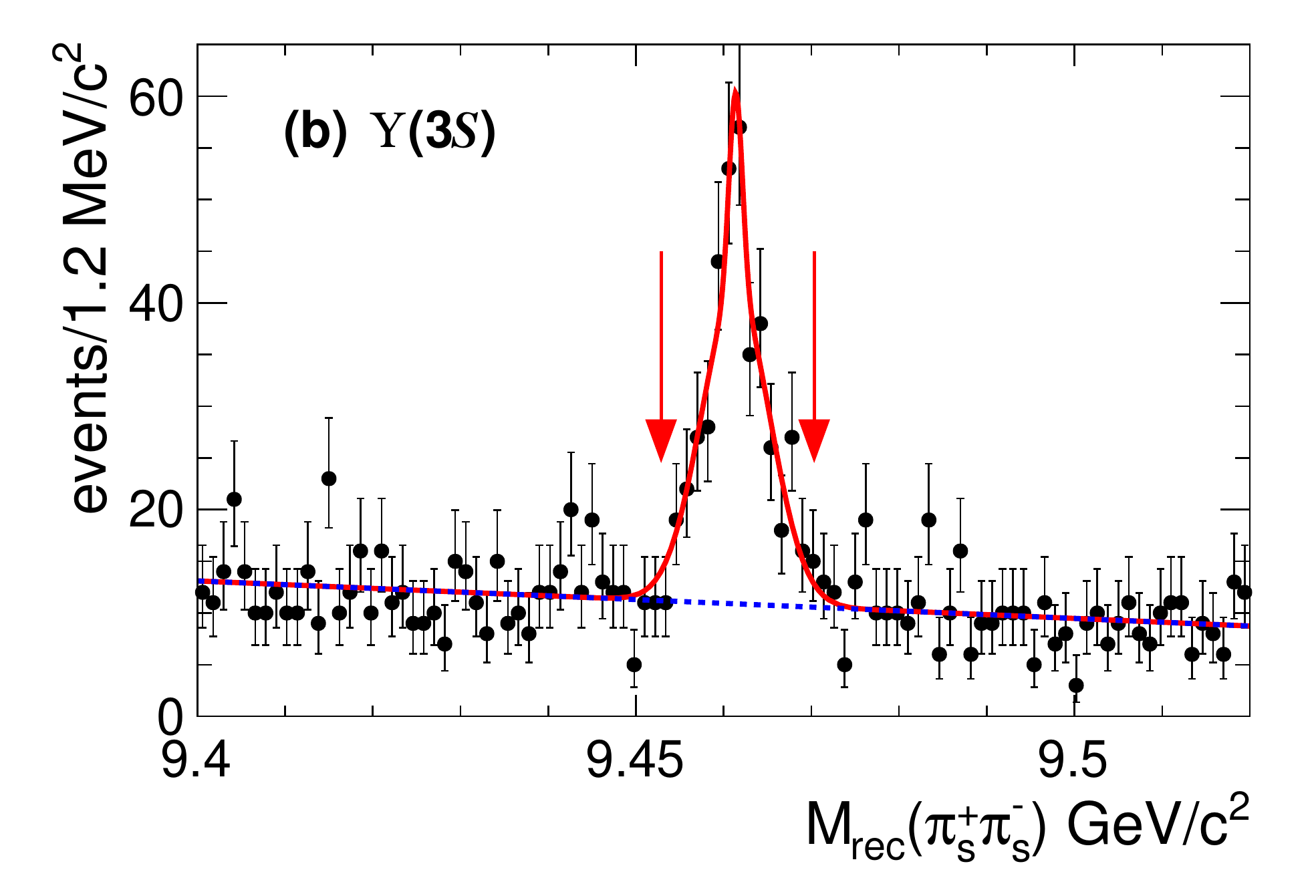}
\caption{Combinatorial recoiling mass \mm to $\pips \pims$ candidates for
 (a) \TwoS and (b) \ThreeS data. The arrows indicate the regions used to select the \OneS signal.}
\label{fig:fig1}
\end{figure}

Events with balanced momentum are required to satisfy energy balance
requirements. For
each combination of $\pips \pims$ candidates, we first require both
particles to be identified loosely as pions and compute the recoiling
mass
\begin{equation}
\mmtwo(\pips \pims) = |p_{\ep} + p_{\en} - p_{\pips} - p_{\pims}|^2,
\end{equation}
where $p$  is the particle four-momentum. The distribution
of $\mmtwo(\pips \pims)$ is expected to peak at the squared \OneS mass for signal events.
Figure~\ref{fig:fig1} shows the combinatorial recoiling mass $\mm(\pips \pims)$
for $\TwoS$ and $\ThreeS$ data, where narrow peaks at the \OneS mass
can be observed.

 We select signal event
candidates by requiring
\begin{equation}
|\mm(\pi^+_s \pi^-_s) - m(\OneS)_f|<2.5 \sigma ,
\label{eq:mrec}
\end{equation}
where $m(\OneS)_f$ indicates the fitted \OneS mass value and $\sigma=2.3$ \mevcc and \mbox{$\sigma$=3.5 \mevcc} for \TwoS and \ThreeS data, respectively.
To reconstruct $\OneS \to \gamma \pip \pim$ or $\OneS \to \gamma \Kp \Km$ decays,
we require a loose identification of both pions or kaons and isolate the two \OneS decay modes by requiring
\begin{equation}
9.1 \ \gevcc < m(\gamma h^+ h^-)< 9.6 \ \gevcc,
\label{eq:mpipig}
\end{equation}
where $h=\pi,K$.

\subsection{Study of the $\pip \pim$ and $\Kp \Km$ mass spectra}

\begin{figure}
  \centering
  \includegraphics[width=7cm]{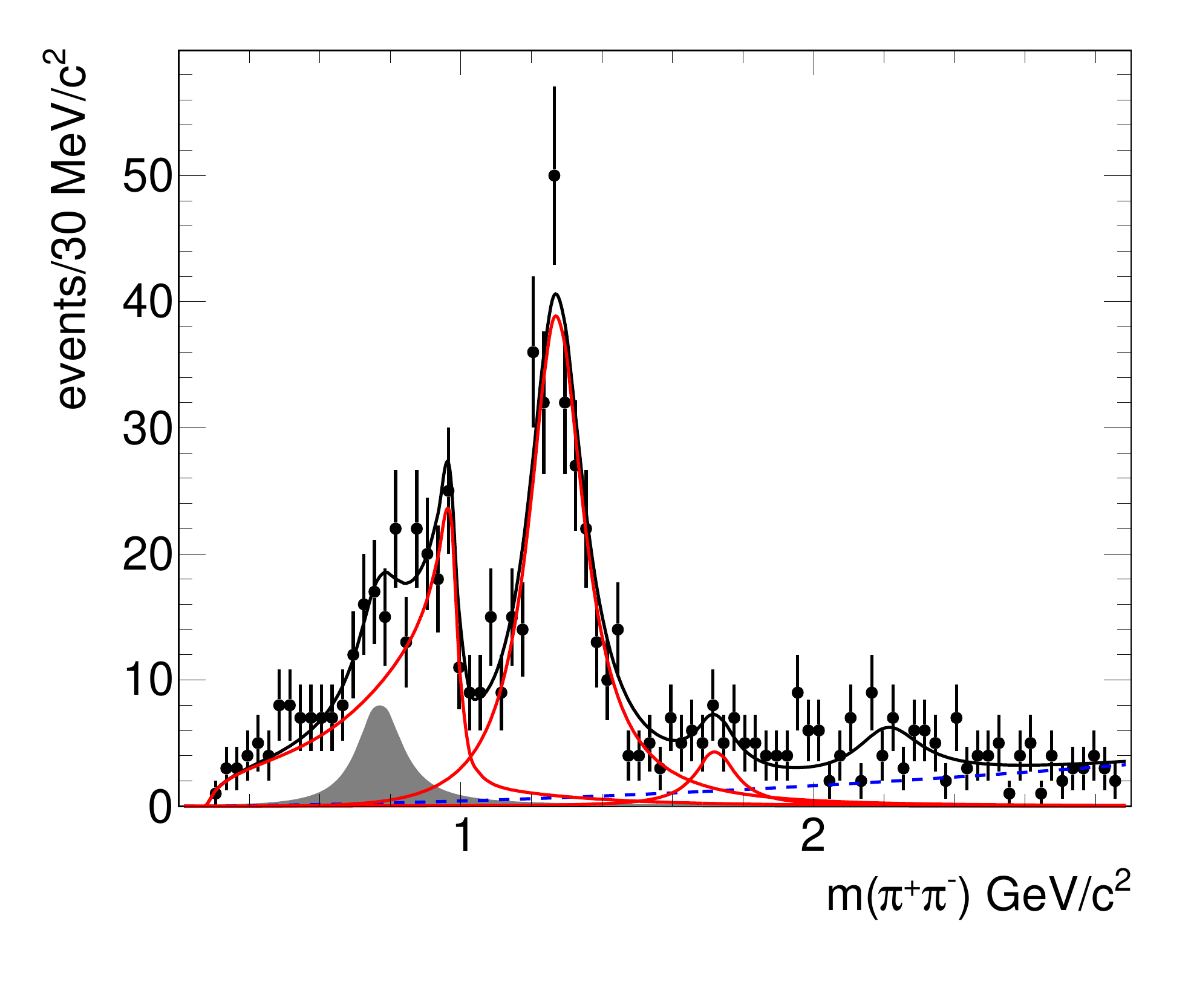}
  \includegraphics[width=7cm]{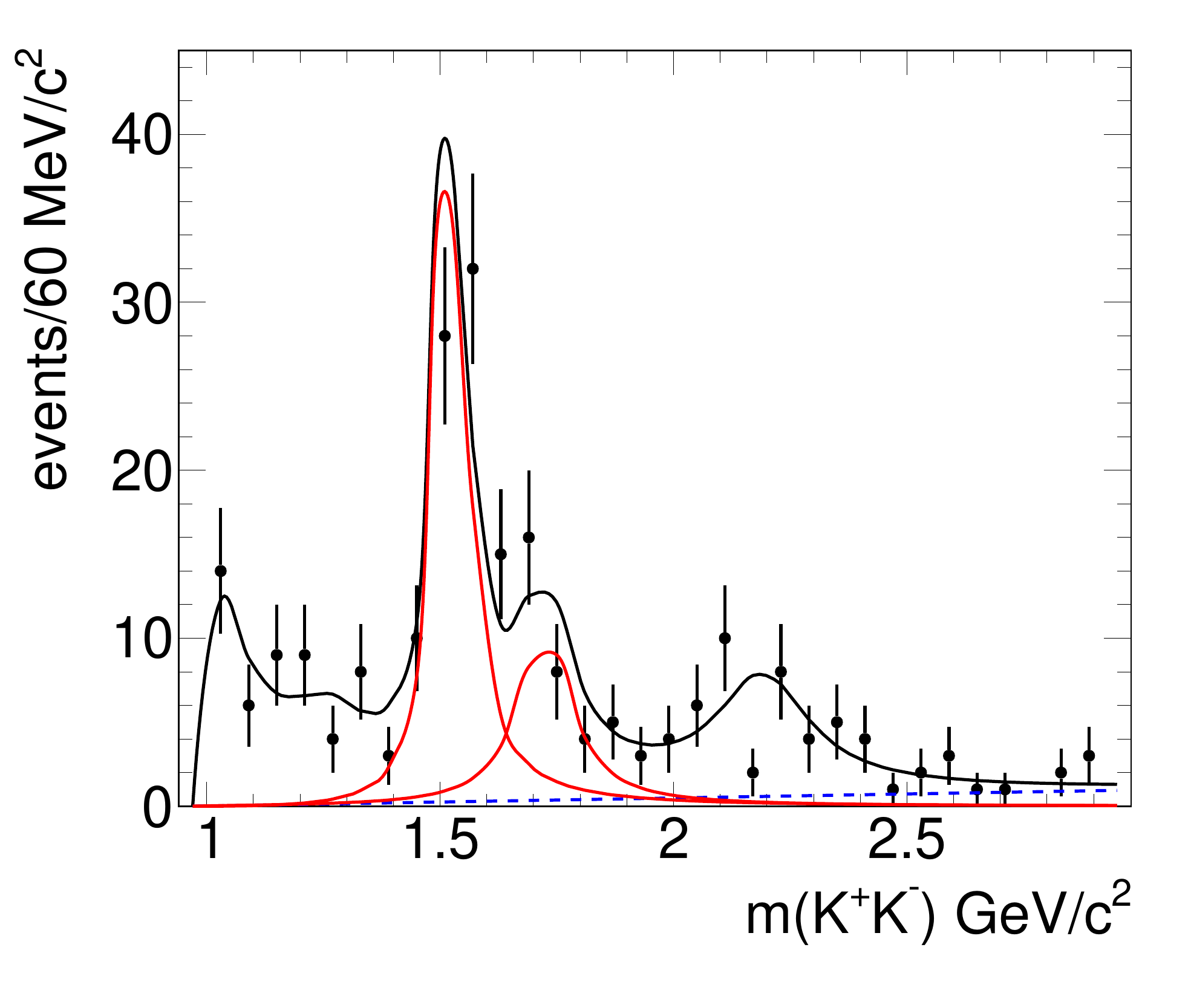}  
  \caption{(Left) $\pip \pim$ mass distribution from $\OneS \to \pip \pim \gamma$ for the combined \TwoS and
  \ThreeS datasets.
  The full (red) curves indicate the $S$-wave, $f_2(1270)$, and $f_0(1710)$ contributions.
  The shaded (gray) area
  represents the estimated $\rho(770)^0$ background.
  (Right) $\Kp \Km$ mass distribution from $\OneS \to \Kp \Km
  \gamma$ for the combined \TwoS and
  \ThreeS datasets. 
  The (red) curves show the contributions from $f_2'(1525)/f_0(1500)$ and $f_0(1710)$.
  Dashed (blue) lines indicate the background contributions.}
\label{fig:fig2}
\end{figure}

The $\pip \pim$ mass spectrum, for $m(\pip \pim)<3.0$ \gevcc\
and summed over the \TwoS and \ThreeS datasets 
with 507 and 277 events, respectively, is shown in Fig.~\ref{fig:fig2}(Left).
The spectrum shows $I=0$, $J^P={\rm even}^{++}$ resonance production, with low backgrounds above 1 \gevcc. We
observe a rapid drop around 1 \gevcc\ characteristic of the presence of the
$f_0(980)$, and a strong $f_2(1270)$ signal. The data also suggest the
presence of additional weaker resonant contributions.

The $\Kp \Km$ mass spectrum, summed over the \TwoS and \ThreeS datasets with
164 and 63 events, respectively, is shown in Fig.~\ref{fig:fig2}(Right) and also shows resonant production, with
low background.
Signals at the positions of $f_2'(1525)/f_0(1500)$ and $f_0(1710)$ can be
observed, with further unresolved structure at higher mass.

We make use of a phenomenological model to extract the different $\OneS \to \gamma R$ branching fractions,
where $R$ is an intermediate resonance.
We perform a simultaneous binned fit to the $\pip \pim$ mass spectra from the \TwoS and \ThreeS datasets.
  We describe the low-mass region (around the $f_0(500)$) using a relativistic $S$-wave Breit-Wigner lineshape having free parameters.
  We describe the $f_0(980)$ using the Flatt\'{e}~\cite{Flatte:1976xu} formalism with parameters fixed to the values from ref.~\cite{WA76:1991kef}.
  The $f_2(1270)$ and $f_0(1710)$ resonances are represented
    by relativistic Breit-Wigner functions with parameters fixed to
    PDG values~\cite{PDG2016}. In the high $\pip \pim$ mass region 
we include a single resonance $f_0(2100)$ having a width fixed to the PDG value
($224 \pm 22$) and unconstrained mass.
For the \ThreeS data we also include $\rho(770)^0$ background with parameters fixed to the PDG values.
The fit is shown in Fig.~\ref{fig:fig2}. It has 16 free parameters and $\chi^2=182$ for ndf=152, corresponding to a $p$-value of 5\%.
We note the observation of a significant $S$-wave in \OneS radiative decays. This observation
was not possible in the study of $\jpsi$ radiative decay to $\pip \pim$ because of the presence of a strong,
irreducible background from $\jpsi \to \pip \pim \piz$~\cite{Mark-III:1986mfz}.
No evidence is found for a $\OneS \to \pip \pim \piz$
decay in present data.
We obtain the following $f_0(500)$ parameters:
\begin{equation}
  m(f_0(500)) = 0.856 \pm 0.086 \ \gevcc, \Gamma(f_0(500))=1.279 \pm 0.324 \ \gev, 
\end{equation}
and $\phi=2.41 \pm 0.43$ rad. The fraction of $S$-wave events associated with the $f_0(500)$ is $(27.7\pm3.1)\%$.

We perform a binned fit to the combined $\Kp \Km$ mass spectrum using the following model.
The $f_0(980)$ is parameterized according to the  Flatt\'{e} formalism.
The $f_2(1270)$, $f_2'(1525)$, $f_0(1500)$, and $f_0(1710)$ resonances are represented by relativistic Breit-Wigner functions with parameters fixed to PDG values. We include an $f_0(2200)$ contribution having parameters fixed to the PDG values.
    The fit shown in Fig.~\ref{fig:fig2}(Right). It has six free parameters and $\chi^2=35$ for ndf=29,
    corresponding to a $p$-value of 20\%.
    The resonances yields and significances are given in  Table~\ref{tab:table1}. Systematic uncertainties are dominated by the PDG uncertainties on resonances
    parameters.
    
   \begin{table}[h]
  \caption{Resonances yields and statistical significances from the
    fits to the $\pi^+ \pi^-$ and $\Kp \Km$ mass spectra for the \TwoS
    and \ThreeS datasets. The symbol $f_J(1500)$ indicates the signal in the 1500 \mevcc\
    mass region.}
   \label{tab:table1}
\begin{center}
\begin{tabular}{lccc}
  \hline
\noalign{\vskip2pt}
Resonances ($\pip \pim$)&  Yield \TwoS &  Yield \ThreeS & Significance\cr
\hline
\noalign{\vskip2pt}
$S$-wave & $ 133 \pm 16 \pm 13$ &  $87 \pm 13$  & 12.8$\sigma$ \cr
$f_2(1270)$ & $255 \pm 19 \pm \al 8$ & $77 \pm 7 \pm 4$ & 15.9$\sigma$\cr
$f_0(1710)$ & $ \al 24 \pm \al 8 \pm \al 6$ & $\al 6  \pm 8 \pm 3$ & \al\all 2.5$\sigma$\cr 
\hline
\noalign{\vskip2pt}
Resonances ($\Kp \Km$)&  Yield \TwoS + \ThreeS &  & Significance \cr
\hline
\noalign{\vskip2pt}
$f_0(980)$ & $47 \pm \al 9 $ &   & 5.6$\sigma$ \cr 
$f_J(1500)$ & $77 \pm 10 \pm 10$ &  & 8.9$\sigma$ \cr
$f_0(1710)$ & $ 36 \pm 9 \pm \al 6 $  & & 4.7$\sigma$ \cr 
\hline
\end{tabular}
\end{center}
   \end{table}

 The efficiency distributions as functions of mass, for the \TwoS/\ThreeS data and for the $\pip \pim \gamma$ and $\Kp \Km \gamma$ final states, are found to have an almost uniform behavior for all the final states.
 We define the helicity angle $\theta_H$ as the angle formed by the $h^+$, in the $h^+ h^-$ rest frame, and the
$\gamma$ in the $h^+ h^- \gamma$ rest frame.
We also define $\theta_{\gamma}$ as the angle formed by the radiative photon in the $h^+ h^- \gamma$ rest frame with respect to the
\OneS direction in the \TwoS/\ThreeS rest frame.
We label with $\epsilon(m,\cos \theta_H)$ the efficiency  computed as a
function of the $h^+ h^-$ effective mass and the helicity angle $\cos \theta_H$. 
We label with $\epsilon(\cos \theta_H,\cos \theta_{\gamma})$ the efficiency computed, for each resonance mass window, as a function of
$\cos \theta_H$ and $\cos \theta_{\gamma}$. This is used to obtain the
efficiency-corrected 
angular distributions and branching fractions for the different
resonances.
To obtain the efficiency correction weight $w_R$ for the resonance $R$ we
divide each event by the efficiency $\epsilon(\cos \theta_H,\cos \theta_{\gamma})$

\begin{equation}
w_R = \frac{\sum_{i=1} ^{N_R}1/\epsilon_i(\cos \theta_H,\cos \theta_{\gamma})}{N_R},
\label{eq:weff}
\end{equation}
\noindent where $N_R$ is the number of events in the resonance mass range.

\subsection{Angular analysis}

\begin{figure}
  \centering
  \includegraphics[width=12cm]{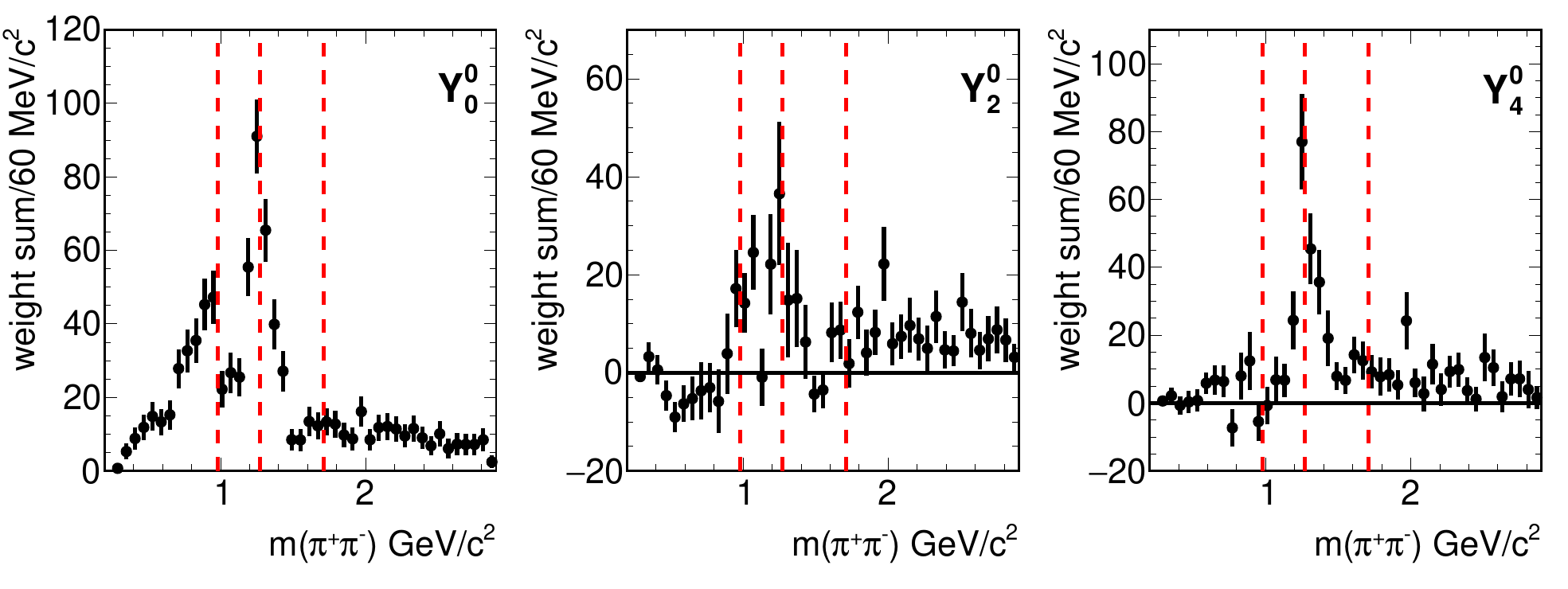}
  \includegraphics[width=12cm]{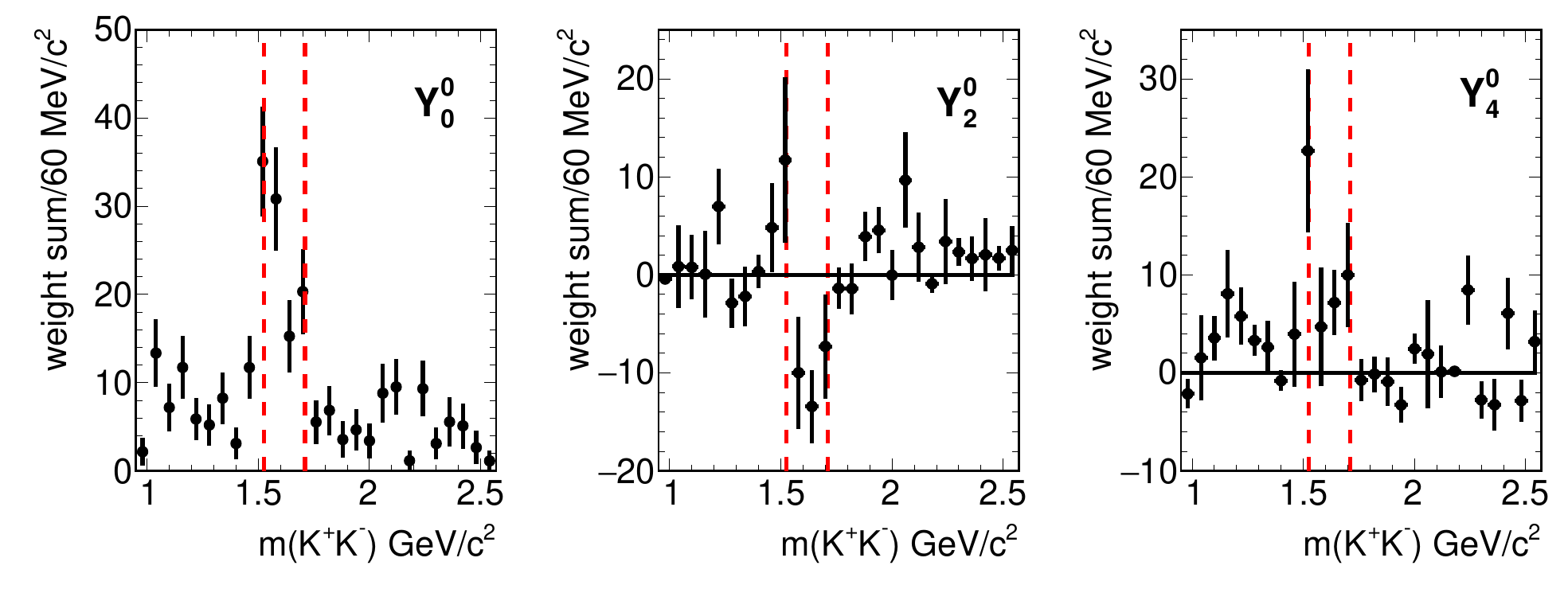}
\caption{The distributions of the most relevant unnormalized $Y^0_L$ moments for $\OneS \to
  \gamma \pip \pim$ (Top) and $\OneS \to
  \gamma \Kp \Km$ (Bottom) as functions of mass.
  The lines indicate the positions of $f_0(980)$, $f_2(1270)$, and $f_0(1710)$ for $\pip \pim$ and $f_2'(1525)$ and $f_0(1710)$ for $\Kp \Km$.}
\label{fig:fig3}
\end{figure}

To obtain information on the angular momentum structure of the $\pip \pim$ and
$\Kp \Km$ systems in $\OneS \to \gamma h^+ h^-$
we study the dependence of the $m(h^+ h^-)$ mass on the helicity angle
$\theta_H$.
A better way to observe angular effects is to 
plot the $\pip \pim$ mass spectrum weighted by the Legendre polynomial
moments, corrected for efficiency and shown in Fig.~\ref{fig:fig3}.
In a simplified environment, the moments are related to the spin 0 ($S$) and \mbox{spin 2 ($D$)} amplitudes by the equations

\begin{equation}
\begin{split}
\sqrt{4 \pi}\langle Y_0^0\rangle & = S^2 + D^2, \\
\sqrt{4 \pi}\langle Y_2^0\rangle & = 2SD\cos\phi_{SD} + 0.639 D^2, \\
\sqrt{4 \pi}\langle Y_4^0\rangle & = 0.857D^2,\\
\end{split}
\label{eq:pwa}
\end{equation}
\noindent
where $\phi_{SD}$ is the relative phase.
Therefore we expect to observe spin 2 resonances in $\langle Y_4^0\rangle$ and
$S/D$ interference in $\langle Y_2^0\rangle$.
The results are shown in Fig.~\ref{fig:fig3}(Top).
We clearly observe the $f_2(1270)$
resonance in $\langle Y_4^0\rangle$ and
a sharp drop in $\langle Y_2^0\rangle$ at the $f_2(1270)$ mass, indicating the
interference effect. 
The distribution of $\langle Y_0^0\rangle$ is just the scaled $\pip \pim$ mass distribution, corrected for efficiency.
Similarly, we plot in Fig.~\ref{fig:fig3}(Bottom) the $\Kp \Km$ mass spectrum weighted by the Legendre polynomial moments, corrected for efficiency.
We observe signals of the $f_2'(1525)$ and $f_0(1710)$ in $\langle Y_4^0\rangle$ and activity due to $S/D$ interference effects in the $\langle Y_2^0\rangle$ moment.

\begin{figure}
  \centering
  \includegraphics[width=12cm]{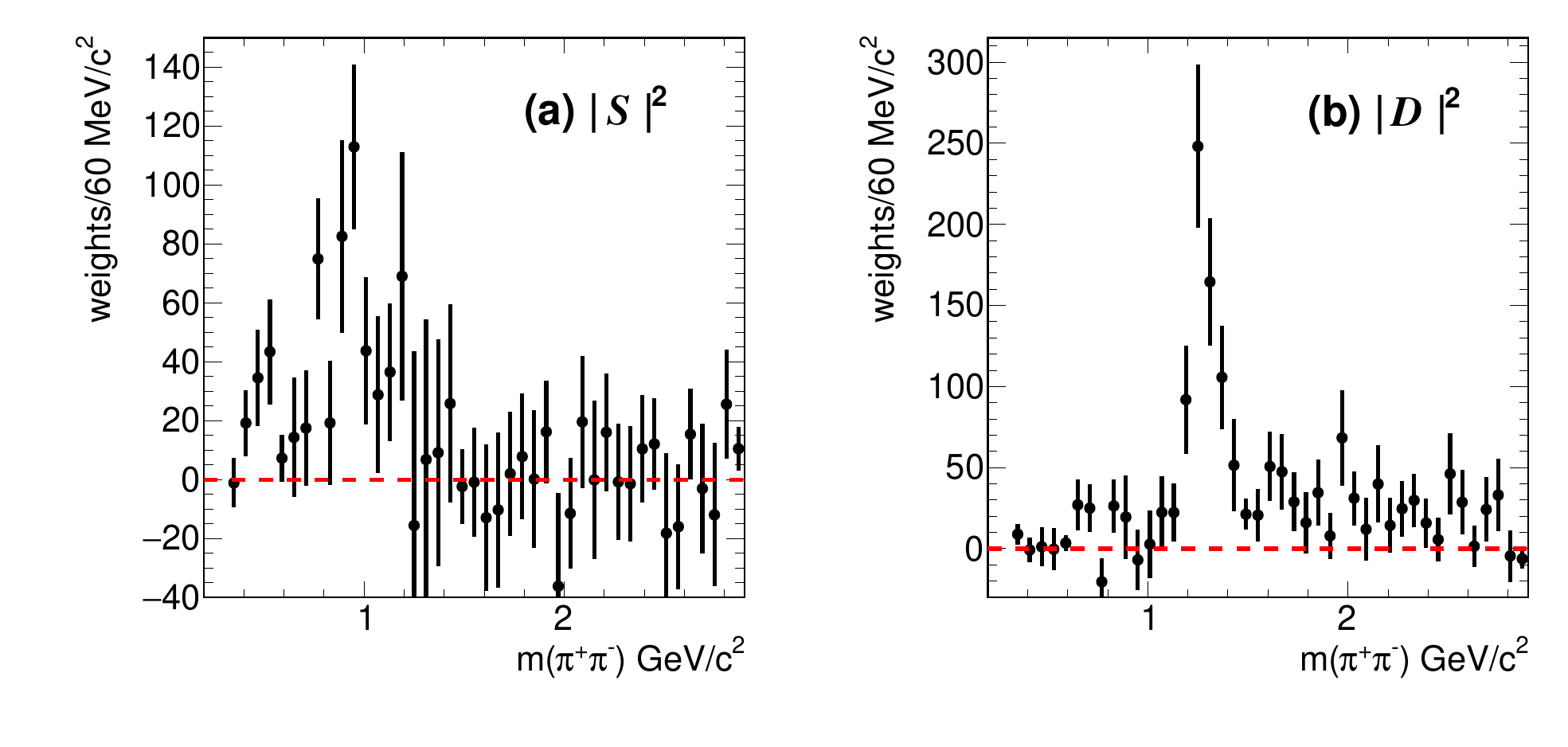}
   \includegraphics[width=12cm]{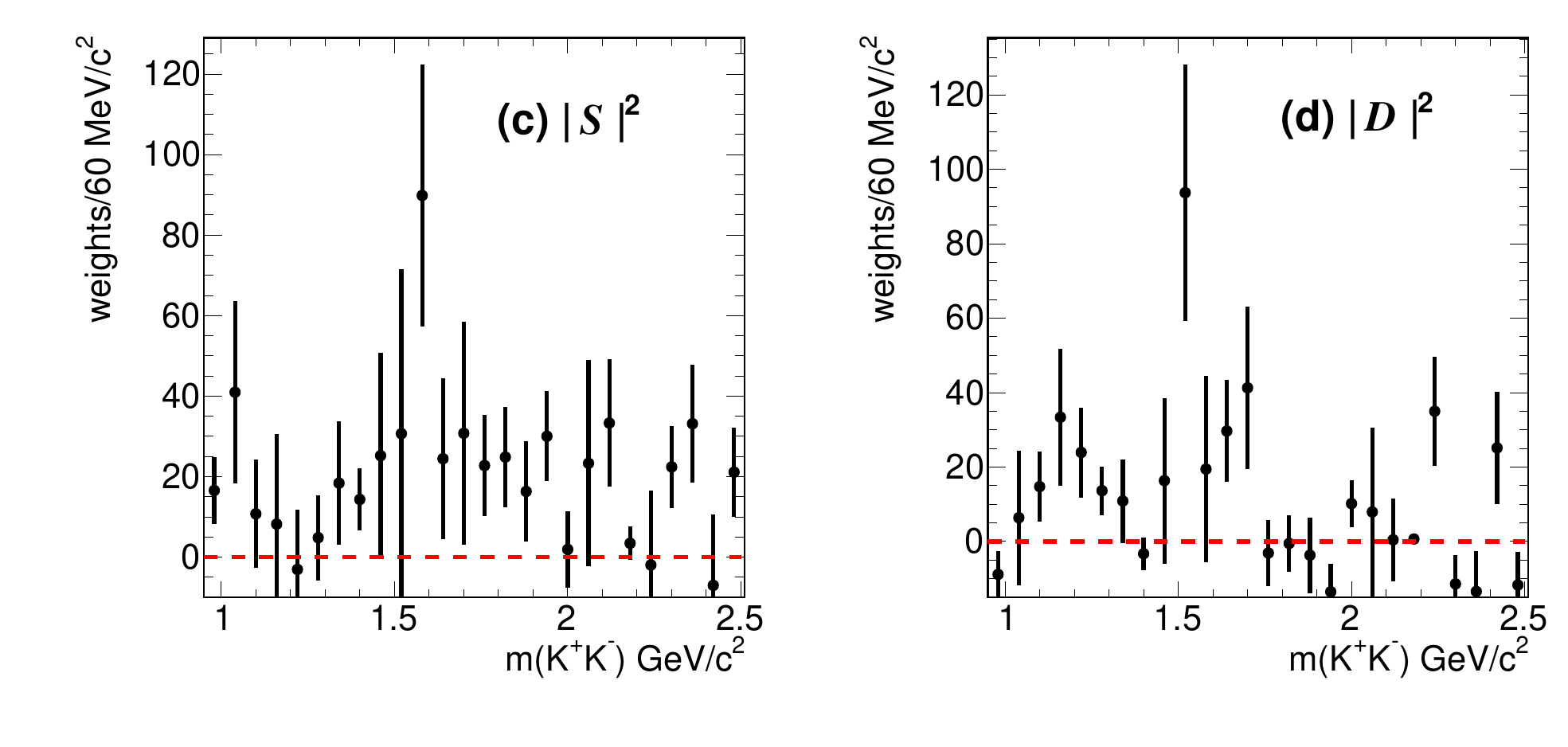}
    \caption{(a)-(c) $S$ and (b)-(d) $D$-wave contributions from the simple PWA of the $\pip \pim$ mass spectrum for the \TwoS data (Left) and of the $K^+ K^-$ mass spectrum for the combined \TwoS/\ThreeS data (Right).}
\label{fig:fig4}
 \end{figure}

Resonance angular distributions in radiative \OneS decays from \TwoS/\ThreeS decays are rather complex (see ref.~\cite{BaBar:2018uqa} for details).
Here we only perform a simplified Partial Wave Analysis (PWA) solving directly the system of Eq.~(\ref{eq:pwa}).
Figure~\ref{fig:fig4}  shows the resulting $S$-wave and $D$-wave contributions to the $\pip \pim$ and $\Kp \Km$
mass spectra, respectively. Due to the presence of background in the threshold region, the $\pip \pim$ analysis is performed only on the \TwoS data.

We note that in the case of the $\pip \pim$ mass spectrum we obtain a good separation between $S$ and $D$-waves, with the presence
of an $f_0(980)$ resonance on top of a broad $f_0(500)$ resonance in the $S$-wave and a clean $f_2(1270)$ in the $D$-wave distribution.
Integrating the $S$-wave amplitude from threshold up to a mass of 1.5 \gevcc, we obtain an integrated, efficiency corrected
yield $N(S{\scriptstyle{-}}{\rm wave}) = 629 \pm 128$.

In the case of the $K^+ K^-$ PWA the structure peaking around 1500 \mevcc\ appears in both $S$ and $D$-waves suggesting the presence
of $f_0(1500)$ and $f_2'(1525)$. In the $f_0(1710)$ mass region there is not enough data to discriminate between the two different
spin assignments. This pattern is similar to that observed in the Dalitz plot analysis of charmless $B \to 3K$ decays~\cite{BaBar:2012iuj}.
Integrating the $S$ and $D$-wave contributions in the $f_2'(1525)/f_0(1500)$ mass region, we obtain a fraction of $S$-wave contribution $f_S(K^+ K^-) = 0.53 \pm  0.10$.

\subsection{\boldmath Measurement of branching fractions}

We determine the branching fraction $\calB(R)$ for the decay of \OneS to photon and resonance $R$ using the expression

\begin{equation}
  \calB(R) = \frac{N_R(\Upsilon(nS)\to \pi^+_s \pi^-_s \OneS(\to R \gamma))}{N(\Upsilon(nS)\to \pi^+_s \pi^-_s \OneS(\to \mup \mun))}\times \calB(\OneS \to \mup \mun),
\label{eq:br}
\end{equation}
\noindent
where $N_R$ indicates the efficiency-corrected yield for the given
resonance. To reduce systematic
uncertainties, we first compute the relative branching fraction to the
reference channel
$\Upsilon(nS)\to \pip \pim \OneS(\to \mup \mun)$, which has the same
number of charged particles as the final states under study. We then
multiply the relative branching fraction by the well-measured branching
fraction $\calB(\OneS \to \mup \mun) =2.48 \pm 0.05$\%~\cite{PDG2016}.

We determine the reference channel corrected yield using the method of
``$B$-counting'', also used to obtain the number of produced \TwoS and \ThreeS~\cite{BaBar:2014omp}. Taking into account the known branching fractions of
$\TwoS/\ThreeS \to \pi^+_s \pi^-_s \OneS$ we obtain
\begin{equation}
  N(\TwoS \to \pi^+_s \pi^-_s \OneS (\to \mup \mun)) = (4.35 \pm 0.12_{\rm sys})\times 10^5
  \label{eq:bc1}
\end{equation}
and
\begin{equation}
  N(\ThreeS \to \pi^+_s \pi^-_s \OneS (\to \mup \mun)) = (1.32 \pm 0.04_{\rm sys}) \times 10^5
  \label{eq:bc2}
\end{equation}
\noindent events.
\noindent
Table~\ref{tab:table2} gives the measured branching fractions.
In all cases we correct the efficiency corrected yields for isospin and for
PDG measured branching fractions~\cite{PDG2016}.
\begin{table}[htb]
  \caption{Measured $\OneS \to \gamma R$ branching fractions.}
   \label{tab:table2}
\begin{center}
\begin{tabular}{lc}
  \hline
\noalign{\vskip2pt}
Resonance & \calB ($10^{-5}$) \cr
\hline
\noalign{\vskip2pt}
$\pi \pi$  $S$-wave & $\al \al \all 4.63 \pm 0.56 \pm 0.48$ \cr
$f_2(1270)$ & $\al\all 10.15 \pm 0.59 \al \all ^{+0.54}_{-0.43}$ \cr
$f_0(1710) \to \pi \pi$ & $\al \al \all 0.79 \pm 0.26 \pm 0.17$ \cr
\hline
$f_J(1500)\to K \bar K$ & $\al \al \all 3.97 \pm 0.52 \pm 0.55$ \cr
$f_2'(1525)$ & $\al \al \all 2.13 \pm 0.28 \pm 0.72$ \cr
$f_0(1500)\to K \bar K$ & $\al \al \all 2.08 \pm 0.27 \pm 0.65$ \cr
\hline
$f_0(1710) \to K \bar K $ & $\al \al \all 2.02 \pm 0.51 \pm 0.35$\cr
\hline
\end{tabular}
\end{center}
\end{table}

We report the first observation of $f_0(1710)$ in \OneS radiative decay with a significance of $5.7\sigma$, combining $\pip \pim$ and $\Kp \Km$ data.
To determine the branching ratio of the $f_0(1710)$ decays to $\pi \pi$ and $K \bar K$, we remove all the systematic uncertainties related to the reference
channels and of the $\gamma$ reconstruction and obtain
\begin{equation}
  \frac{\calB(f_0(1710)\to \pi\pi)}{\calB(f_0(1710)\to K \bar K)} 
   =  0.64 \pm 0.27_{\rm stat} \pm 0.18_{\rm sys},
\end{equation}
in agreement with the world average value of $0.41^{+0.11}_{-0.17}$~\cite{PDG2016}.

\section{Dalitz plot analysis of $\eta_c$ three-body decays}

The results presented here are based on the full data set collected
with the \babar\ detector using an integrated luminosity of 519~\invfb recorded at
center-of-mass energies at and near the $\Upsilon (nS)$ ($n=2,3,4$)
resonances. 
 In the present analysis, we consider the three-body \etac decays to \etaprkk, \etaprpipi, and \etapipi ,  
using two-photon interactions, $\epem\to\epem\gamma^*\gamma^*\to\epem\etac$, where $\gamma^*$ indicate the intermediate quasi-real virtual photons.

\subsection{Study of $\gamma \gamma \to \etapr h^+ h^-$ and $\gamma \gamma \to \eta \pip \pim$}

We first study the reactions
\begin{equation}
  \gamma \gamma \to \etapr h^+ h^-,
  \label{eq:etaphh}
\end{equation}
where $h^+h^-$ indicates a $\pip \pim$ or $\Kp \Km$ system. 
The \etapr is reconstructed in the two decay modes $\etapr \to \rho^0 \gamma$, $\rho^0 \to \pip \pim$, and $\etapr \to \eta \pip \pim$, $\eta \to \gamma \gamma$.
We define \pt\ as the magnitude of the transverse momentum of the $\etapr h^+h^-$ system, in the \epem\ rest frame, with respect to the beam axis.
Well reconstructed two-photon events with quasi-real photons are expected to have low values of \pt .
For the selection of the $\etapr \pip \pim$ final state, we require all four charged tracks to be positively identified as pions.
For the \etaprkk final state, 
we require the two charged tracks assigned to the \etapr decay to be positively identified as pions and 
the other two to be positively identified as kaons.
We require $\pt<0.05$~\gevc\ and $\pt<0.15$~\gevc, for the $\etapr \to \rho^0 \gamma$ and $\etapr \to \eta \pip \pim$, respectively.
We discriminate against Initial State Radiation (ISR) events $\epem \to \gamma_{ISR} h^+ h^-$, by requiring the recoil mass $\mm\equiv(p_{\epem}-p_{\rm rec})^2 > 2$~GeV$^2$/$c^4$,
where $p_{\epem}$ is the four-momentum of the initial state \epem\, 
and $p_{\rm rec}$ is the reconstructed four-momentum of the candidate $\etapr(\eta) h^+h^-$ system.

\begin{figure}
  \begin{center}
  \includegraphics[width=7.0cm]{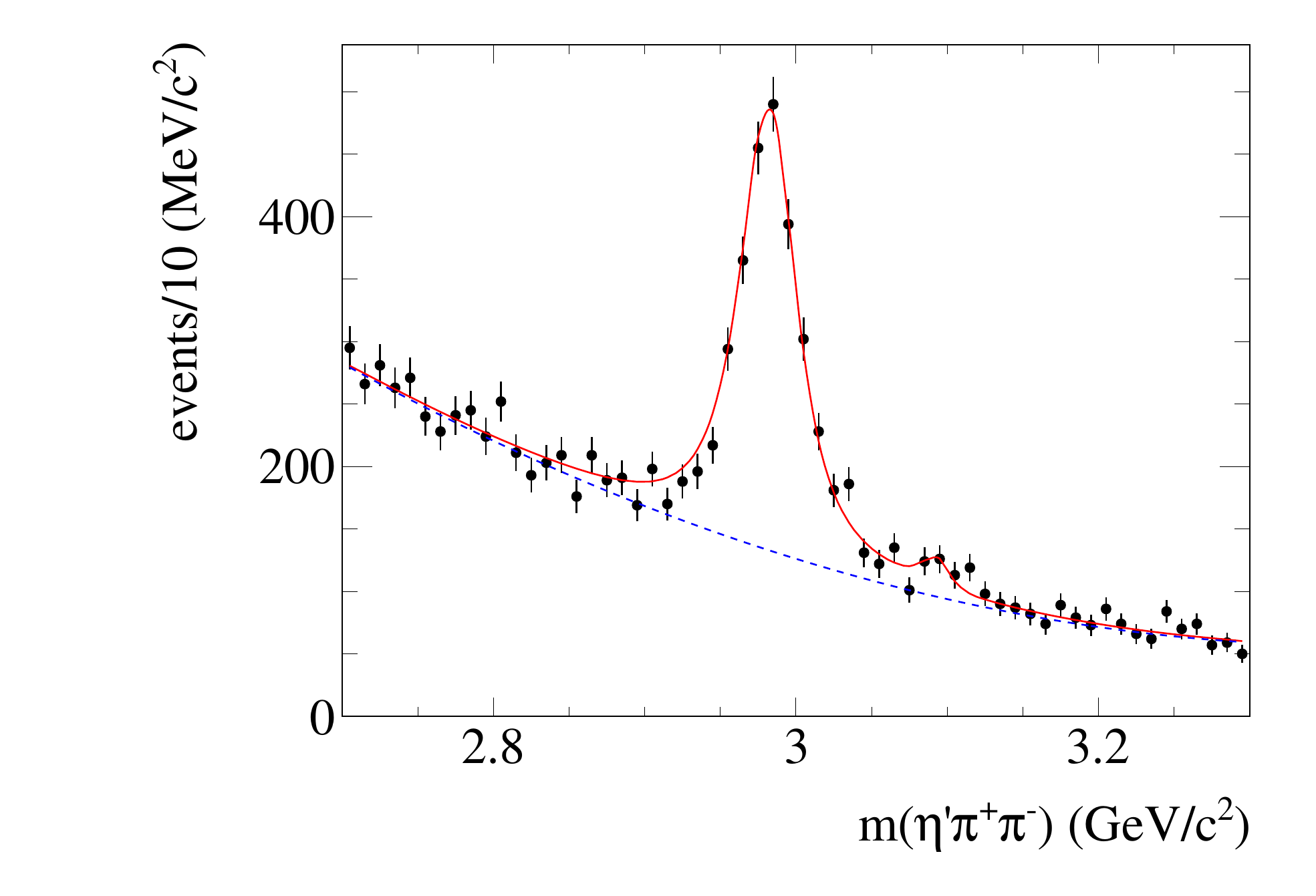}
  \includegraphics[width=7.0cm]{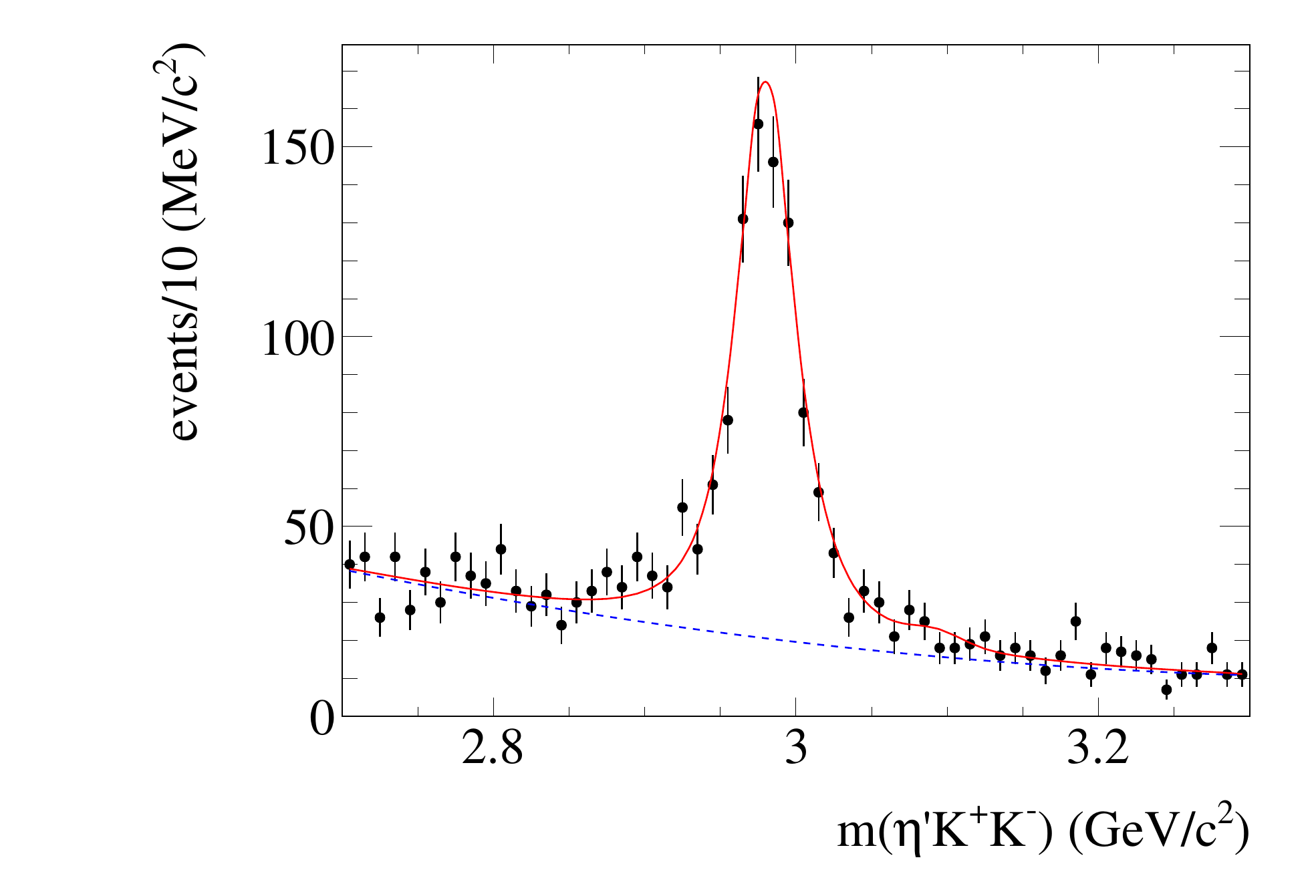}    
\caption{Invariant-mass distributions of selected (Left) \etaprpipi and (Right) \etaprkk candidates summed over the $\etapr \to \rho^0 \gamma$ and $\etapr \to \eta \pip \pim$ decay modes.}
\label{fig:fig5}
\end{center}
\end{figure}

The \etaprpipi and \etaprkk mass spectra, summed over the $\etapr \to \rho^0 \gamma$ and $\etapr \to \eta \pip \pim$ decay modes are shown in fig.~\ref{fig:fig5}, where prominent \etac signals can be observed.
In particular, fig.~\ref{fig:fig5}(Right) reports the first observation of the decay $\etac \to \etapr \Kp \Km$.

\begin{figure}
  \begin{center}
    \includegraphics[width=7.0cm]{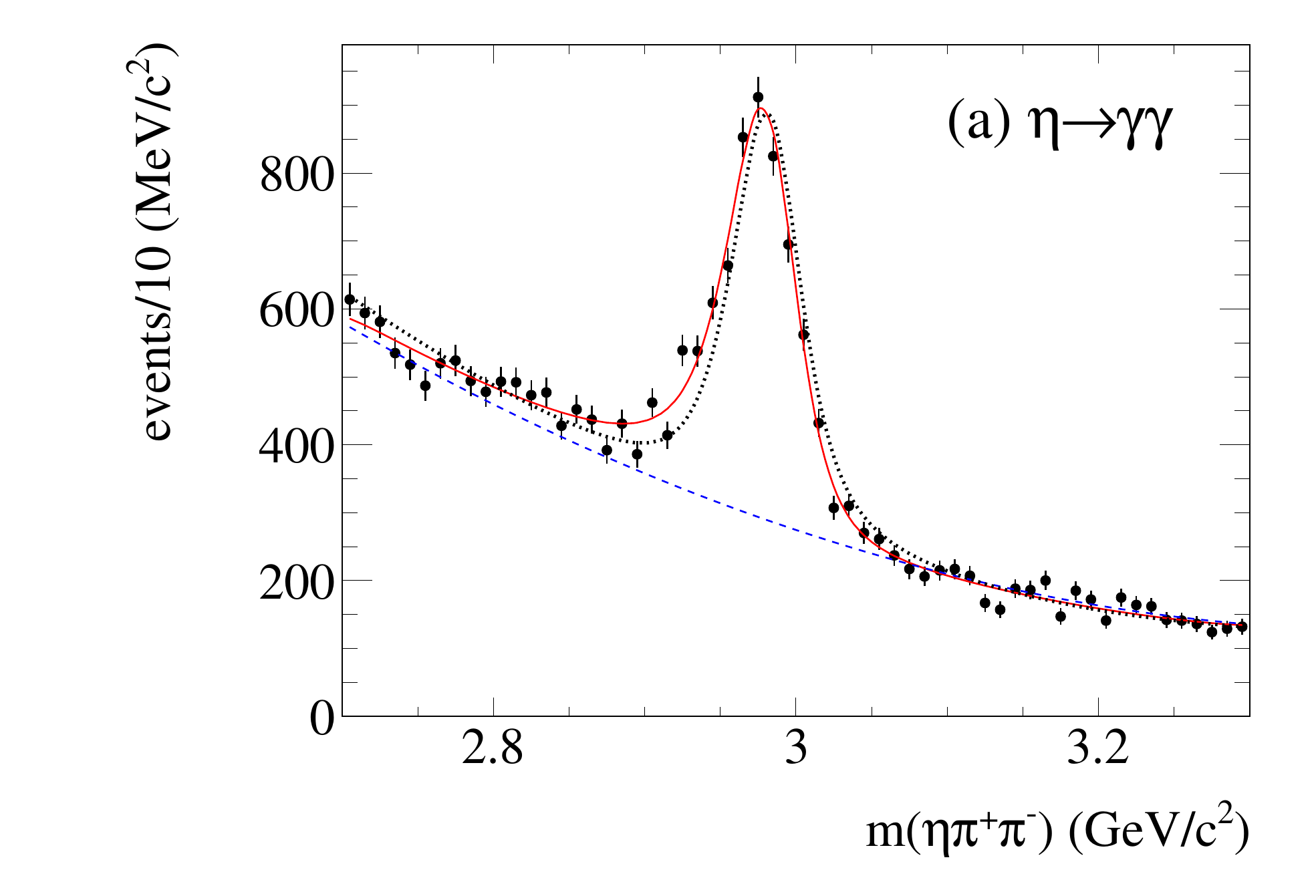}
    \includegraphics[width=7.0cm]{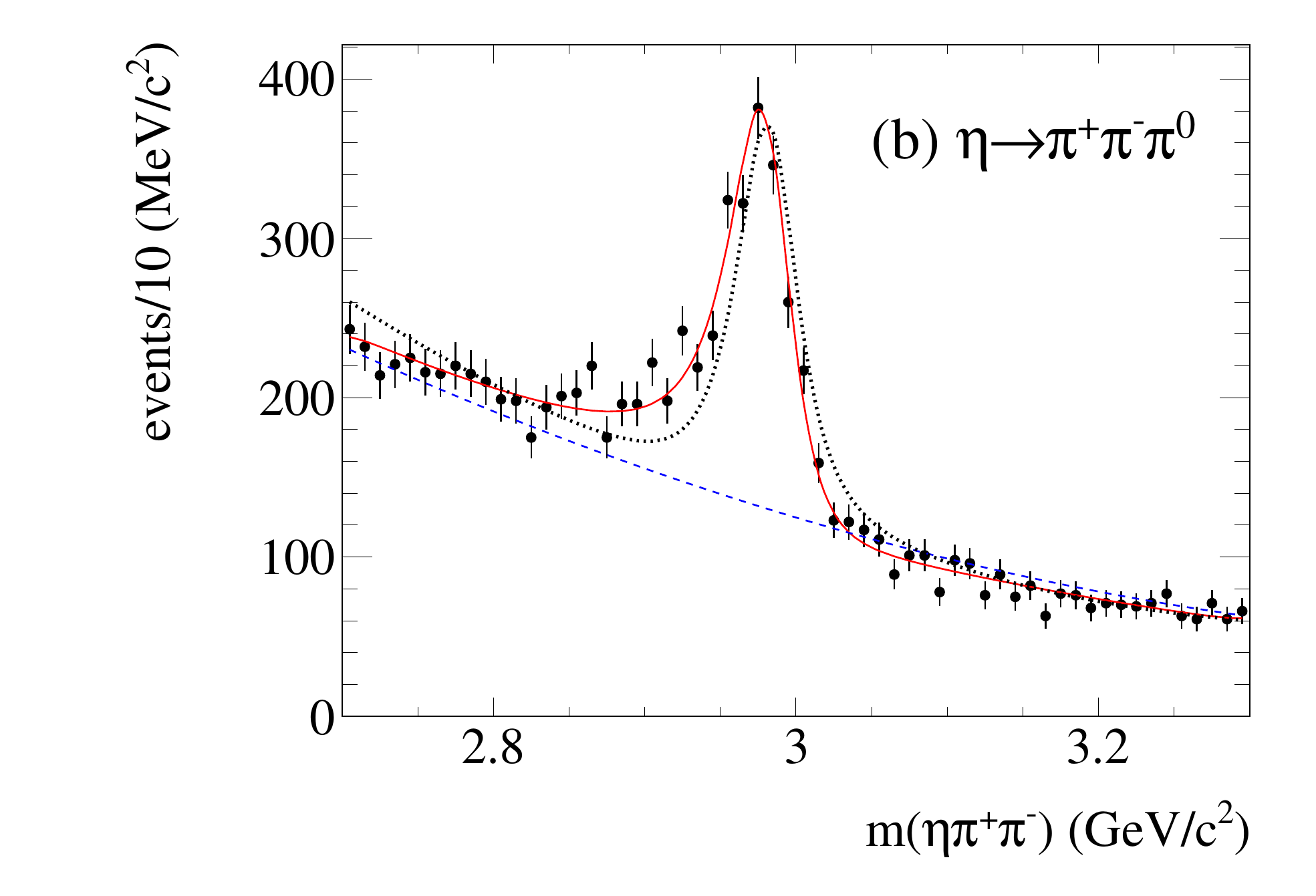}
    \caption{ Invariant-mass spectra for selected $\eta \pip \pim$ candidate events with (a) $\eta \to \gamma \gamma$ and (b) $\eta \to \pip \pim \piz$.
    The solid (red) lines represent the fits including interference described in the text. The dashed (blue) line represents the fitted non-resonant components. The dotted lines represent the fits without interference.}
\label{fig:fig6}
\end{center}
\end{figure}

We also study the reaction
\begin{equation}
  \gamma \gamma \to \eta \pip \pim, 
  \label{eq:etapipi}
\end{equation}
where $\eta \to \gamma \gamma$ and $\eta \to \pip \pim \piz$.
In this case the two-photon reaction is selected by requiring $\pt<0.1$ \gevc\ for both $\eta$ decay modes. The corresponding $\eta \pip \pim$ mass spectra are shown in fig.~\ref{fig:fig6}, where prominent \etac signals can be observed.

To compute the reconstruction and selection efficiency, MC signal events are generated using a detailed detector simulation in which the \etac\ mesons decay uniformly in phase space.
These simulated events are reconstructed and analyzed in the same manner as data. 
We define the helicity angle $\theta_H$ as the angle formed by the $h^+$ (where $h=\pi,K$), in the $h^+ h^-$ rest frame, and the \etapr ($\eta$) direction in the $h^+ h^- \etapr$ ($h^+ h^- \eta$) rest frame.
To smoothen statistical fluctuations, the efficiency maps are parameterized using Legendre polynomials up to $L=12$
as functions of $\cos \theta_H$ in intervals of $m(h^+ h^-)$ and then interpolated linearly between adjacent mass intervals.

\subsection{Yields and branching fractions}

We fit the invariant-mass distributions to obtain the numbers of selected \etac\ events, $N_{\etapr K^+ K^-}$,  $N_{\etapr \pi^+ \pi^-}$, and $N_{\eta \pi^+ \pi^-}$, for each \etapr\ or $\eta$ decay mode. 
We then use the \etaprkk and \etaprpipi yields to compute the ratio of branching fractions for \etac to the \etaprkk and \etaprpipi final states.  This ratio is computed as
\begin{equation}
\calR = \frac{\BR(\etac \to \etapr \Kp \Km)}{\BR(\etac \to \etapr \pip \pim)} 
  = \frac{N_{\etapr K^+ K^-}}{N_{\etapr \pi^+ \pi^-}}\frac{\epsilon_{\etapr \pi^+ \pi^-}}{\epsilon_{\etapr K^+ K^-}}
\end{equation}
for each \etapr decay mode, 
where $\epsilon_{\etapr K^+ K^-}$ and $\epsilon_{\etapr \pi^+ \pi^-}$ are the corresponding efficiencies.
We determine $N_{K^+K^-\etapr}$ and $N_{\pi^+ \pi^- \etapr}$ from \etac decays by performing binned $\chi^2$ fits to the \etaprkk and \etaprpipi invariant-mass spectra, in the 2.7-3.3 \gevcc\ mass region, separately for the two \etapr decay modes. In these fits, the \etac signal contribution is described by a simple Breit-Wigner (BW) function convolved with a fixed resolution function (described by the sum of a Gaussian and Crystal Ball functions), with \etac parameters fixed to PDG values~\cite{PDG2020}.
An additional BW function is used to describe the residual background from ISR $J/\psi$ events, 
and the remaining background is parameterized by a $2^{nd}$ order polynomial.
The fitted $\etapr h^+ h^-$ invariant-mass spectra are shown in Fig.~\ref{fig:fig5} summed over the two \etapr decay modes.

We estimate $\epsilon_{\etapr K^+ K^-}$ and $\epsilon_{\etapr \pi^+ \pi^-}$ for the \etac signals using the \mbox{2-D} efficiency functions described above.
Each event is first weighted by $1/\epsilon(m,\cos \theta_H)$. 
Since the backgrounds below the \etac signals have different distributions in the Dalitz plot, 
we perform a sideband subtraction by assigning an additional weight of $+1$ to events in the \etac signal region, defined as the (2.93-3.03) \gevcc\ mass region, and a
weight $-1$ to events in the sideband regions, (2.77-2.87)~\gevcc\ and (3.09-3.19)~\gevcc.
The two evaluations of the branching fractions, for the two \etapr decay mode, are in good agreement and give an average value of
      \begin{equation}
        \frac{\calB(\etac \to \etapr \Kp \Km)}{\calB(\etac \to \etapr \pip \pim)} = 0.644 \pm 0.039_{\rm stat} \pm 0.032_{\rm sys}.
        \label{eq:br_etapr}
      \end{equation}
For the decay $\etac \to \eta \pip \pim$ the fits without
interference do not describe the data well for either $\eta$ decay mode.
Leaving free the \etac parameters, the fits return masses shifted down by $\approx10$ \mevcc\ with respect to PDG averages.

We test the possibility of interference effects of the \etac with each non-resonant two-photon process~\cite{Belle:2012uhr},
modifying the fitting function by defining
\begin{equation}
      f(m) = |A_{\rm nres}|^2+|A_{\etac}|^2+c\cdot 2Re(A_{\rm nres} A^*_{\etac}),
      \label{eq:int}
\end{equation}
where $A_{\rm nres}$ is the non-resonant amplitude with $|A_{\rm nres}|^2$ described by a $2^{nd}$ order polynomial;
the coherence factor $c$ is the fraction of the non-resonant events that are true two-photon production of the same final state;
the resonant contribution is described by
\mbox{$A_{\etac}=\alpha\cdot BW(m) \cdot \exp(i\phi)$}, where $BW(m)$ is a simple Breit-Wigner with parameters fixed to PDG values; and $\alpha$, $\phi$, and $c$ are free parameters.
The fitted invariant-mass spectra are shown in Fig.~\ref{fig:fig6}, where reasonable descriptions of the data are evident.
As a comparison we also show the fit the two mass spectra with no interference and fixed \etac parameters and
obtain the dotted lines distributions shown in Fig.~\ref{fig:fig6}.

\subsection{Dalitz plot analysis}

We perform Dalitz plot analyses of the \etaprpipi, \etaprkk, and \etapipi systems in the \etac mass region using unbinned maximum likelihood fits. Amplitudes are parameterized as described in Ref.~\cite{Asner:2003gh}.
They include a relativistic Breit-Wigner function having a variable width modulated by the Blatt-Weisskopf~\cite{Blatt:1952ije} spin form factors
and the relevant spin-angular information.
We first fit the two \etac\ sidebands separately, using an incoherent sum of amplitudes.
To model the background composition in the \etac\ signal region, we take a weighted average of the two fitted fractional contributions, and normalize using the results from the fit to the \etac\ signal region.

\subsubsection{Dalitz plot analysis of $\etac \to \etapr \Kp \Km$.}

\begin{figure}
  \begin{center}
    \includegraphics[width=8.5cm]{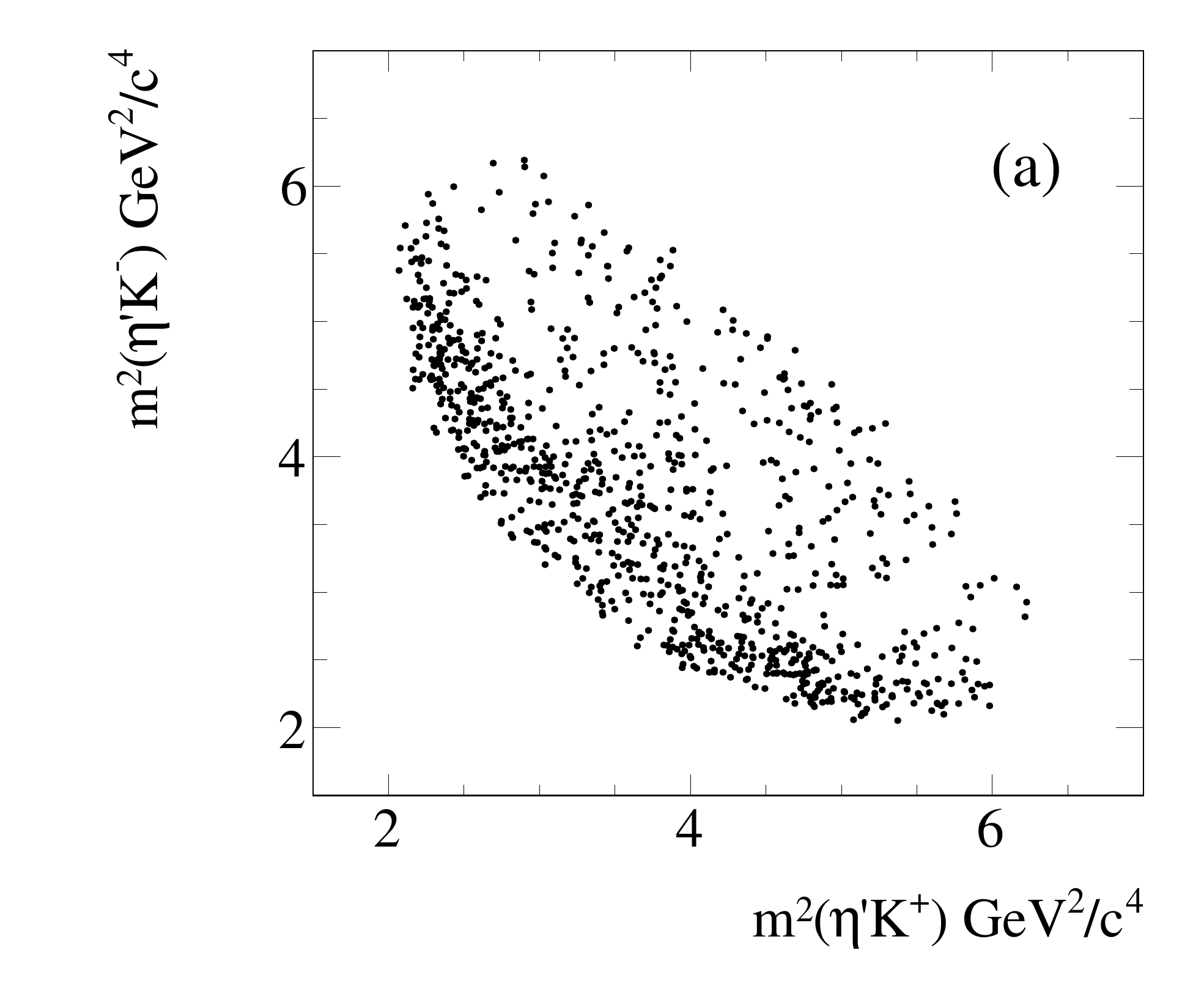}
    \includegraphics[width=14cm]{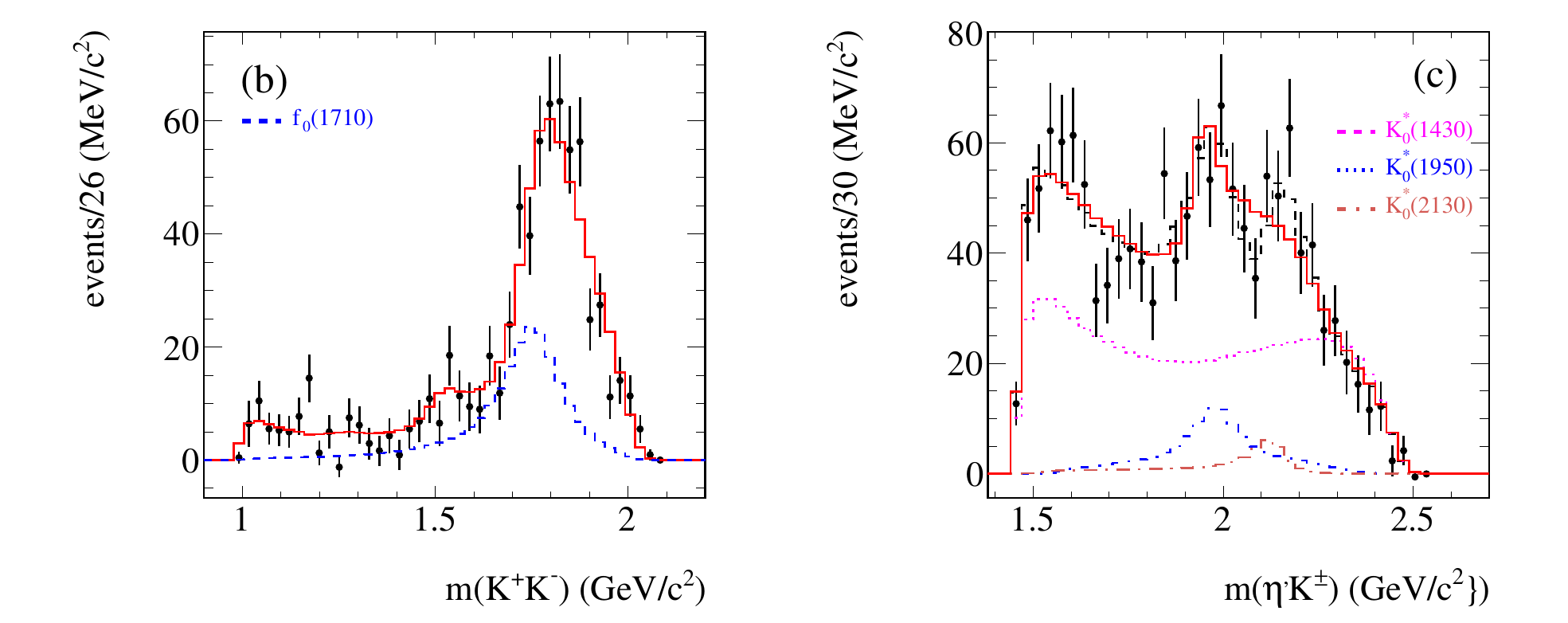}    
\caption{(a) Dalitz plot for selected $\etac \to \etapr \Kp \Km$ candidates in the \etac\ signal region, summed over the two \etapr decay modes. Linear-scale mass projections (b) $m(\Kp \Km)$ and (c) $m(\etapr K^{\pm})$, after subtraction of the background.
  The solid (red) histograms represent the results of the fit described in the text.
  The (black) dashed line in (c) shows the solution which include the presence of $K^*_0(2130)$.
The other histograms display the contributions from each of the listed components.}      
\label{fig:fig7}
\end{center}
\end{figure}

\begin{figure}
  \begin{center}
    \includegraphics[width=14cm]{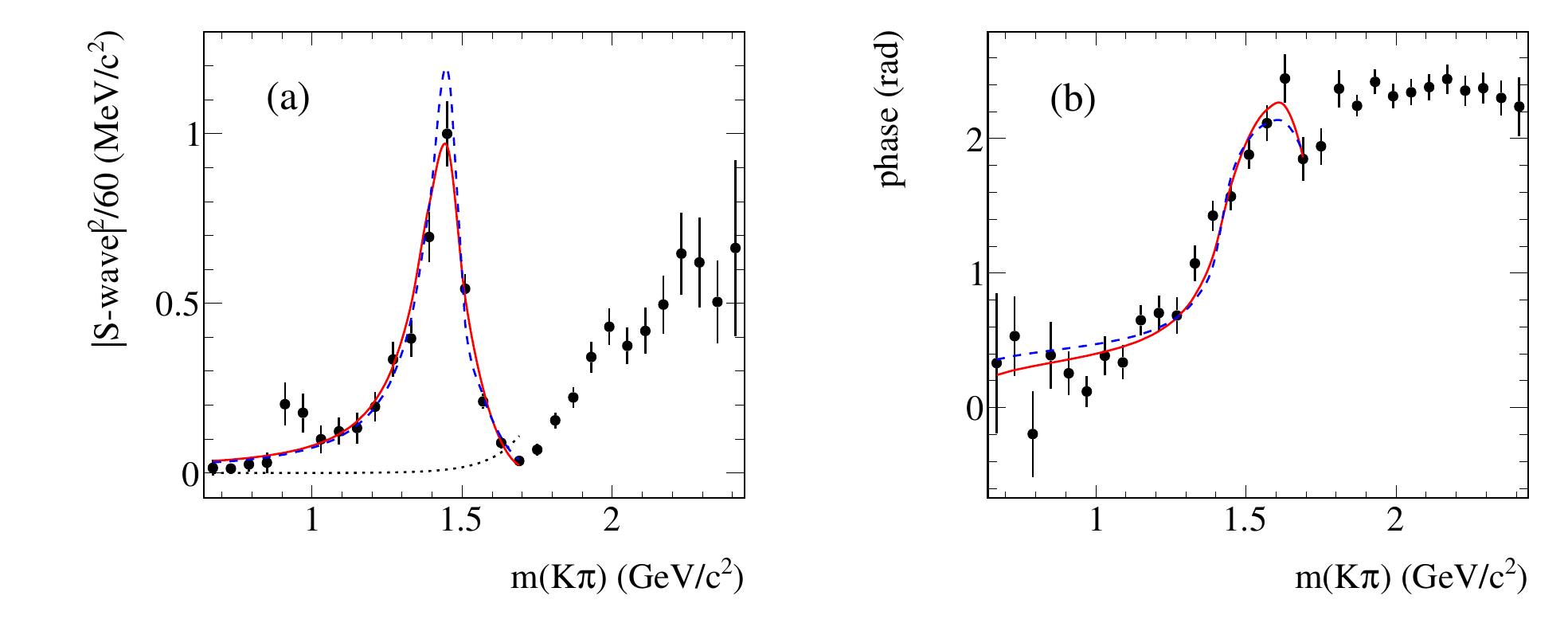}
  \caption{The (a) squared modulus and (b) phase of the $K \pi$ $S$-wave averaged over the $\eta_c \to \KS K \pi$ and
    $\eta_c \to \Kp \Km \piz$ from the \babar~\cite{BaBar:2015kii} QMI analysis. 
    The full (red) lines represent the result from the fit with free $g^2_{K\etapr}$ and $g^2_{K\pi}$ parameters. The dashed (blue) lines
    represent the result from the fit with a fixed $g^2_{K \etapr}/g^2_{K \pi}$ ratio~\cite{BaBar:2021fkz}.
    The dotted (black) line in (a) represents the empirical background contribution.
}
\label{fig:fig8}
\end{center}
\end{figure}

      Figure~\ref{fig:fig7}(a) shows the Dalitz plot for the selected $\etac \to \etapr \Kp \Km$ candidates in the data, for the two \etapr decay modes combined (930 events).
We observe that this \etac decay mode is dominated by a diagonal band
on the low mass side of the Dalitz plot. The $m(\Kp \Km)$ spectrum shows a large structure in the region of the $f_0(1710)$ resonance. The combined
$m(\etapr K^{\pm})$ invariant-mass spectrum shows a structure at threshold due to the $K^*_0(1430)$ accompanied by weaker resonant structures.
The $K^*_0(1430)$ is a relatively broad resonance decaying to $K \pi$,
$K \eta$, and $K \etapr$. The measured $K \eta$ relative branching fraction is $\frac{\BR(K^*_0(1430) \to K \eta)}{\BR(K^*_0(1430) \to K \pi)} = 0.092 \pm 0.025^{+0.010}_{-0.025}$~\cite{BaBar:2014asx}, while the $K \etapr$ has only been observed in Ref.~\cite{BESIII:2014dlb}. To describe the $K^*_0(1430)$ lineshape in the $K \etapr$ projection, we model it using a simplified coupled-channel Breit-Wigner function, which ignores the small $K \eta$ contribution. We parameterize the $K^*_0(1430)$ signal as

\begin{equation}
        BW(m) = \frac{1}{m_0^2 - m^2 - i(\rho_1(m)g^2_{K \pi} + \rho_2(m)g^2_{K \etapr})},
        \label{eq:ch}
\end{equation}
\noindent
where $m_0$ is the resonance mass, $g_{K \pi}$ and $g_{K \etapr}$ are the couplings to the $K \pi$ and $K \etapr$ final states, and \mbox{$\rho_j(m)=2P/m$} are the respective Lorentz-invariant
phase-space factors, with $P$ the decay particle momentum in the $K^*_0(1430)$ rest frame. 
The values of $m_0$ and the $g_{Kj}$ couplings cannot be derived from the $K\etapr$ system only,
and therefore we make use of the $K \pi$ $S$-wave measurement from \babar~\cite{BaBar:2015kii}. 
We average the reported quasi model-independent (QMI) measurements of the $K\pi$ $S$-wave from $\eta_c \to \KS K \pi$ and $\eta_c \to \Kp \Km \piz$ decays, and obtain the modulus squared of the amplitude and the phase shown in Fig.~\ref{fig:fig8}.

We perform a simultaneous binned $\chi^2$ fit to the $K \pi$ $S$-wave amplitude and phase from threshold up to 1.72~\gevcc. 
We model the $K \pi$ $S$-wave in this region as\\
  \mbox{$S$-${\rm wave}(m) = B(m) + c \cdot BW_{K \pi}(m) e^{i \phi}$},
 where $BW_{K \pi}(m)$ is given by Eq.(~\ref{eq:ch}), $B(m)$ is an empirical background term, parameterized as
$B(m) = \rho_1(m)e^{-\alpha m}$,
      and $c$, $\phi$, and $\alpha$ are free parameters.
      The results of the fit are shown in Fig.~\ref{fig:fig8} as the solid (red) lines.
      
      We perform a Dalitz plot analysis of the $\etac \to \etapr \Kp \Km$ decay channel by using the $\etapr f_0(1710)$
      intermediate state as the reference amplitude.
      The projections of the fit result are shown in Fig.~\ref{fig:fig7}(b)-(c), along with the largest signal components.
      We measure the $f_0(1710)$ parameters, listed in Table~\ref{tab:table3}.
      We also measure the parameters of the $K^*_0(1950)$ (see Table~\ref{tab:table3}) for which there is only one previous measurement from the LASS collaboration~\cite{Aston:1987ir}.
      For the $K^*_0(1430)$ resonance we make use of an iterative procedure, combining the results of the present Dalitz plot analysis with previous measurements (see ref.~\cite{BaBar:2021fkz} for details). The results from the Dalitz analysis are given in Table~\ref{tab:table4}.
      \begin{table}[h] 
        \caption{\small Resonance parameters from the Dalitz plot analyses of $\eta_c \to \etapr \Kp \Km$, $\eta_c \to \etapr \pip \pim$, and $\eta_c \to \eta \pip \pim$. Significances are computed using the Wilks theorem~\cite{wilks} and 
do not include systematic uncertainties.
        }
        \centering
 {\small
        \begin{tabular}{lcccc}
    \hline \\ [-2.3ex]
Resonance & Mass (\mevcc) & $g_{K \pi}^2$ (\gevccc) &  $g_{K \etapr}^2$ (\gevccc) & \cr
\hline \\ [-2.3ex]
& &  $\eta_c \to \etapr \Kp \Km$   & &\cr
\hline \\ [-2.3ex]
\hline \\ [-2.3ex]
$K^*_0(1430)$ & & & \cr
\hline \\ [-2.3ex]
$\etac \to K \bar K \pi$ & \almm\almm\alm\aln$1447 \pm 8$ &  \alm\alm$0.414 \pm 0.026$ & $0.197 \pm 0.105$ &\cr
fixed $\frac{g^2_{\etapr K}}{g^2_{\pi K}}$ & \almm\almm\aln$1453 \pm 22$ &  \alm\alm$0.462 \pm 0.036$ &   &\cr
\\ [-2.3ex]
\hline \\ [-2.3ex]
Resonance & Mass (\mevcc) & $\Gamma$ (\mev) &   & significance (n$\sigma)$\cr
\hline \\ [-2.3ex]
$f_0(1710)$ & $1757 \pm 24 \pm 9$ & $175 \pm 23 \pm 4$ & & 11.4 \cr
(a) $K^*_0(1950)$ &  $1942 \pm 22 \pm 5$ & \al $80 \pm 32 \pm 20$ & & 3.3\cr
\hline \\ [-2.3ex]
(b) $K^*_0(1950)$ &  $1979 \pm 26 \pm 3$ & \al $144 \pm 44 \pm 21$ & & 4.3\cr
$K^*_0(2130)$ &  $2128 \pm 31 \pm 9$ & \al $95 \pm 42 \pm 76$ & & 2.7\cr
\hline \\ [-2.3ex]
\hline \\ [-2.3ex]
& &  $\eta_c \to \etapr \pip \pim$  & & \cr
\hline \\ [-2.3ex]
\hline \\ [-2.3ex]
$f_0(500)$ &  \alm\alm$953 \pm 90$ &  \almm\alm\alm\aln$335 \pm 81$ & &\cr
$f_2(1430)$ & $1440 \pm 11 \pm 3$ &  $46 \pm 15 \pm 5$ & & 4.4 \cr
$f_0(2100)$ & \al$2116 \pm 27 \pm 17$ & $289 \pm 34 \pm 15$ & & 10\cr
\hline \\ [-2.3ex]
\hline \\ [-2.3ex]
&  & $\eta_c \to \eta \pip \pim$  & &\cr
\hline \\ [-2.3ex]
\hline \\ [-2.3ex]
$a_0(1700)$ & \alm$1704 \pm 5 \pm 2$ & $110 \pm 15 \pm 11$ & & 8\cr
\hline \\ [-2.3ex]
\hline \\ [-2.3ex]          
  \end{tabular}
 }
\label{tab:table3}
      \end{table}

\begin{table}[h] 
        \caption{Fractions and relative phases from the Dalitz plot analysis of $\eta_c \to \etapr \Kp \Km$.
        }
        \centering
  \begin{tabular}{lcccc}
\hline \\ [-2.3ex]
Intermediate state & fraction (\%) & phase (rad)\cr
\hline \\ [-2.3ex]
$f_0(1710) \etapr$ & $29.5 \pm 4.7 \pm 1.6$ & 0. \cr
$K^*_0(1430)^+ K^-$ & $53.9 \pm 7.2 \pm 2.0$   & \al$0.61 \pm 0.13 \pm 0.45$ \cr
$K^*_0(1950)^+ K^-$ & \al $2.4 \pm 1.2 \pm 0.4$ & \al$0.46 \pm 0.29 \pm 0.50$ \cr
$f_0(1500) \etapr$ & \al$0.8 \pm 1.0 \pm 0.3$ & \al$0.32 \pm 0.54 \pm 0.10$ \cr
$f_0(980) \etapr$ & \al$4.7 \pm 2.7 \pm 0.4$ & \aln$-0.74 \pm 0.55 \pm 0.05$ \cr
$f_2(1270) \etapr$ & \al\all$2.9 \pm 1.5 \pm 0.1$l & \al\al$2.9 \pm 0.38 \pm 0.09$ \cr
\hline \\ [-2.3ex]
sum & $94.3 \pm  9.3 \pm 2.6$ & \cr
$p$-value & 18\% & \cr
\hline \\ [-2.3ex]
  \end{tabular}
\label{tab:table4}
      \end{table}
An inspection of Fig.~\ref{fig:fig7}(c) suggests an additional enhancement in the $m(\etapr K^{\pm})$ around a mass of $\approx 2100$ \mevcc.
We explore this possibility adding, in the Dalitz plot analysis, an additional scalar resonance in this mass region with free parameters.
The presence of this additional resonance also affects the parameters of the $K^*_0(1950)$ which are also left free in the fit.
The results from this solution are listed in Table~\ref{tab:table4}, labelled as solution (b). 
A comparison between the two fits on the $m(\etapr K^{\pm})$ projection is shown in Fig.~\ref{fig:fig7}(c). However, an application of the Wilks theorem
for the individual significances of the $K^*_0(1950$ and $K^*_0(2130)$ in this new fit, obtain values of 4.3$\sigma$ and 2.7$\sigma$, respectively.
Since the local significance of the $K_0^*(2130)$ is less than
$3\sigma$, we do not consider the presence of this contribution in the reference fit.

      \subsubsection{Dalitz plot analysis of $\etac \to \etapr \pip \pim$.}
      
\begin{figure}
  \begin{center}
    \includegraphics[width=8.5cm]{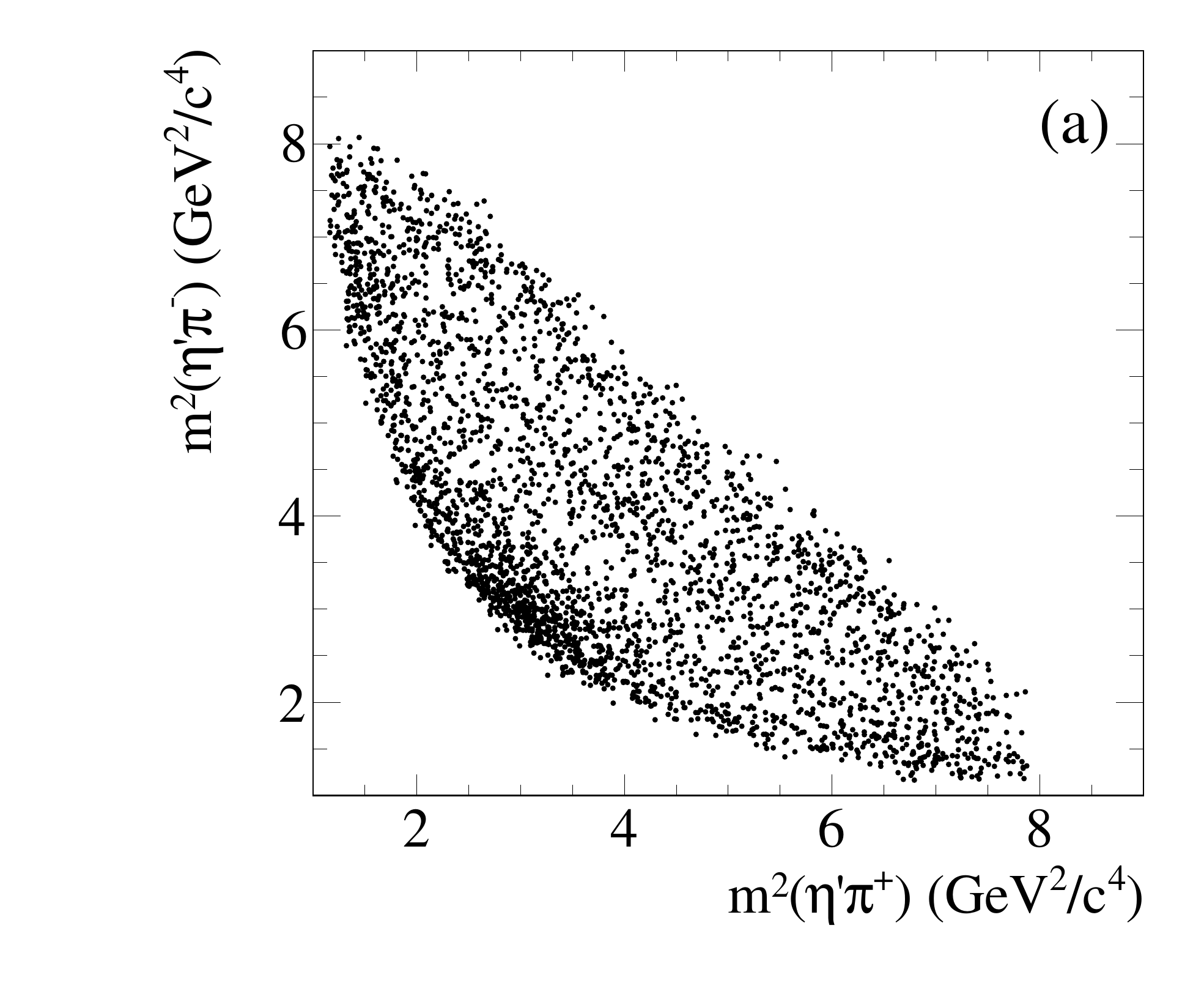}
    \includegraphics[width=14cm]{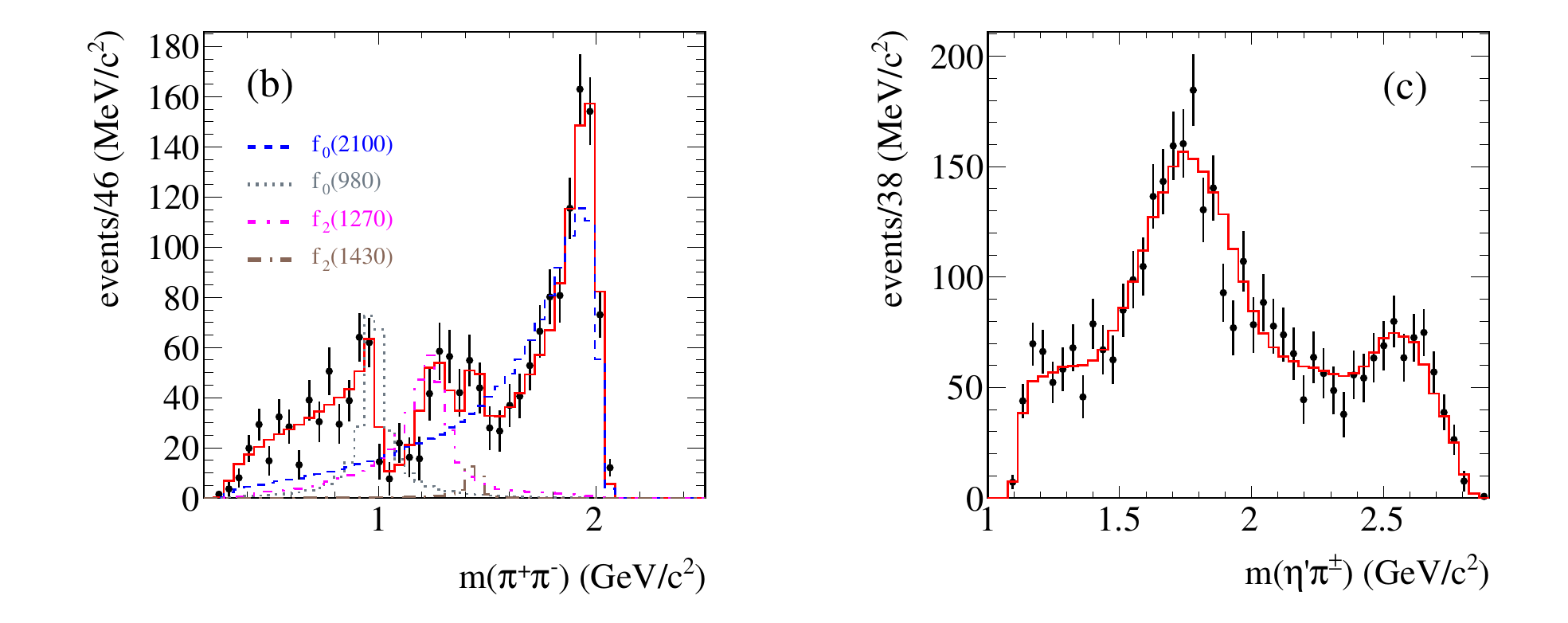}
  \caption{(a) Dalitz plot for selected $\etac \to \etapr \pip \pim$ candidates in the \etac\ signal region, summed over the two \etapr decay modes. Linear-scale mass projections (b) $m(\pip \pim)$ and (c) $m(\etapr\pipm)$, after subtraction of the background.
  The solid (red) histograms represent the results of the fit described in the text, and the other histograms display the contributions from each of the listed components.
  }
\label{fig:fig9}
\end{center}
\end{figure}

Figure~\ref{fig:fig9}(a) shows the Dalitz plot for the selected $\etac \to \etapr \pip \pim$ candidates in the data, in the \etac\ signal region, for the two \etapr decay modes combined (3122 events), and Figs.~\ref{fig:fig9}(b)-(c) show the
two background subtracted projections in linear mass scale. 
We observe several diagonal bands in the Dalitz plot, in particular at the lower-left edge. 
There are corresponding structures in the $m(\pip \pim)$ spectrum, including peaks attributable to the $f_0(980)$ and $f_2(1270)$ resonances, and a large structure at high $\pip \pim$ mass. 
A candidate for the large structure in the high $\pip \pim$ mass region is the $f_0(2100)$ resonance, observed in radiative $J/\psi$ decay to $\gamma \eta \eta$~\cite{Ablikim:2013hq}.
We take $f_0(2100)\etapr$ as the reference contribution, and perform a Dalitz plot analysis whose results are given in Table~\ref{tab:table5}. We leave free the $f_0(2100)$ resonance parameters and obtain the values reported in Table~\ref{tab:table4} with a significance of $10\sigma$.
To describe the small enhancement around 1.43 \gevcc , we test both spin-2 and spin-0 hypotheses with free resonance parameters; 
we obtain $\Delta(-2\log \calL)=2.4$ in favor of the spin-2 hypothesis, so we attribute this signal to the $f_2(1430)$ resonance, and report the fitted parameter values in Table~\ref{tab:table4}. 
We test the significance of this signal by removing it from the list of the resonances, obtaining $\Delta(-2\log \calL)=23.8$ and a significance of $4.4 \sigma$.
      \begin{table}
        \caption{\small Fractions and relative phases from the Dalitz plot analysis of $\eta_c \to \etapr \pip \pim$.
        }
        \centering
  \begin{tabular}{lcc}
\hline \\ [-2.3ex]
Intermediate state & fraction (\%) & phase (rad)\cr
\hline \\ [-2.3ex]
$f_0(2100) \etapr$ & $74.9 \pm 7.5 \pm 3.6$ & 0.  \cr
$f_0(500) \etapr$ & \al$4.3 \pm 2.3 \pm 0.7$ & $-5.89 \pm 0.24 \pm 0.10$ \cr
$f_0(980) \etapr$ & $16.1 \pm 2.4 \pm 0.5$ & $-5.31 \pm 0.16 \pm 0.04$  \cr
$f_2(1270) \etapr$ & $22.1 \pm 2.9 \pm 2.4$ & $-3.60 \pm 0.16 \pm 0.03$  \cr
$f_2(1430) \etapr$ & \al$1.9 \pm 0.7 \pm 0.1$ & $-2.45 \pm 0.32 \pm 0.11$  \cr
$a_2(1710) \pi$ & \al$3.2 \pm 1.9 \pm 0.5$ & $-0.75 \pm 0.27 \pm 0.11$ \cr
$a_0(1950) \pi$    & \al$2.5 \pm 1.1 \pm 0.1$& $-0.02 \pm 0.32 \pm 0.06$ \cr
$f_2(1800) \etapr$ & \al$5.3 \pm 2.2 \pm 1.4$ & \al\all$0.67 \pm 0.24 \pm 0.08$  \cr
\hline \\ [-2.3ex]
sum & \alm$130.5 \pm 9.5 \pm 4.7$ & \cr
$p$-value & 20\% & \cr
\hline \\ [-2.3ex]
  \end{tabular}
\label{tab:table5}
      \end{table}

\subsubsection{Dalitz plot analysis of $\etac \to \eta \pip \pim$.}

\begin{figure}
  \begin{center}
    \includegraphics[width=8.5cm]{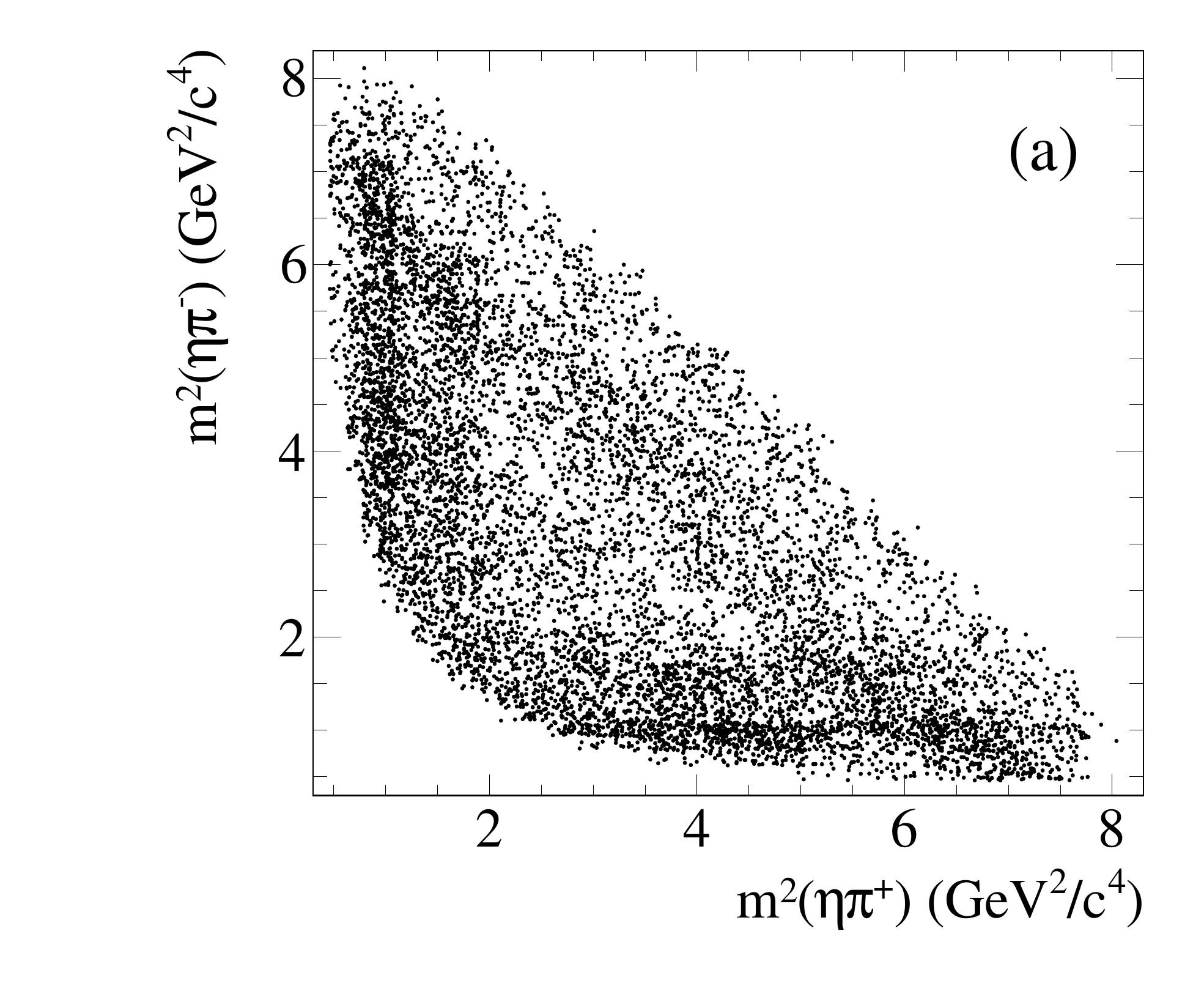}
    \includegraphics[width=14cm]{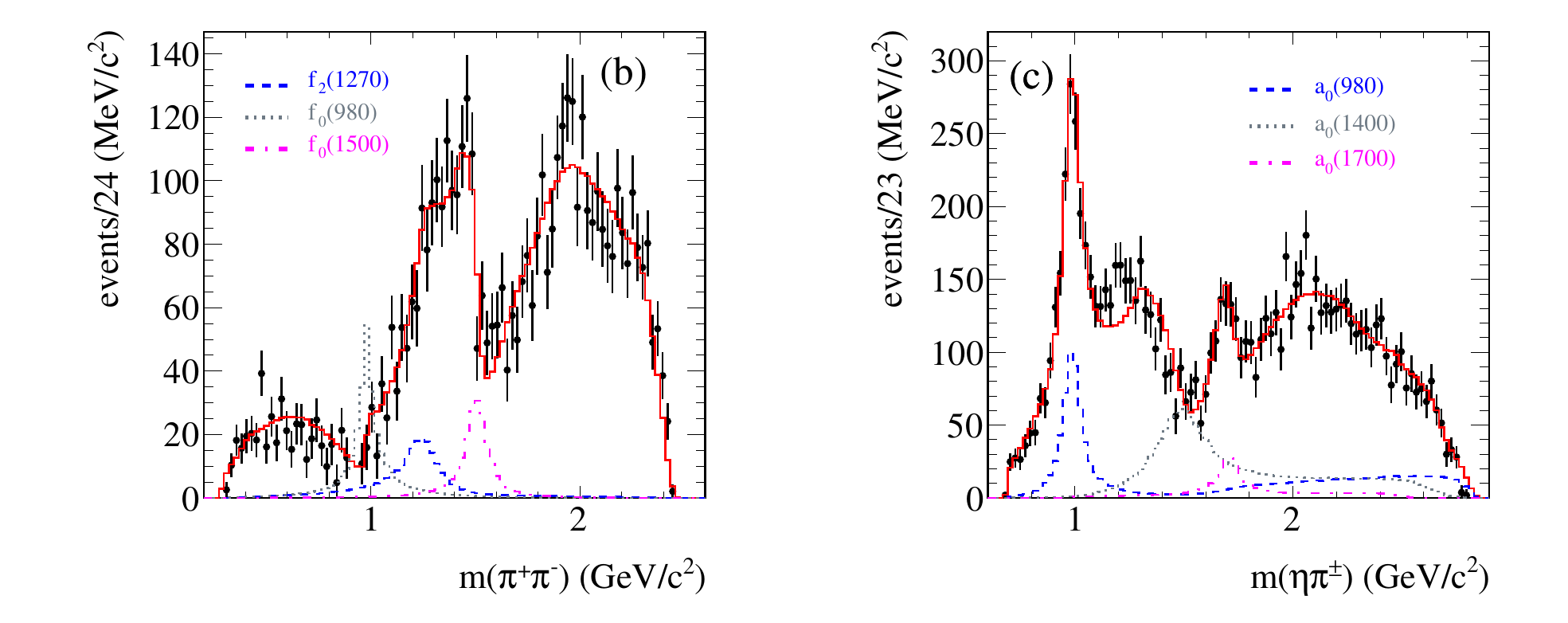}
  \caption{(a) Dalitz plot for selected $\etac \to \eta \pip \pim$ candidates in the \etac\ signal region, summed over the two $\eta$ decay modes. Linear-scale mass projections (b) $m(\pip \pim)$ and (c) $m(\eta \pipm)$, after subtraction of the background.
  The solid (red) histograms represent the results of the fit described in the text, and the other histograms display the contributions from each of the listed components.
  }
\label{fig:fig10}
\end{center}
\end{figure}

Figure~\ref{fig:fig10}(a) shows the Dalitz plot for the selected $\etac \to \eta \pip \pim$ candidates in the data, in the \etac\ signal region, for the two $\eta$ decay modes combined (9303 events), and Figs.~\ref{fig:fig10}(b)-(c) show two background subtracted linear-mass projections. 
We observe that the Dalitz plot is dominated by horizontal and vertical bands due to the $a_0(980)$ and diagonal bands in the $\pip \pim$ final state corresponding to 
$f_0(500)$, $f_0(980)$, and $f_2(1270)$ resonances.
We take $a_0(980)^+ \pim$ as the reference contribution, and perform a Dalitz plot analysis as described above.
The resulting list of contributions to this \etac decay mode is given in Table~\ref{tab:table6}, together with fitted fractions and relative phases.
\begin{table} 
     \caption{\small  Fractions and relative phases from the Dalitz plot analysis of $\eta_c \to \eta \pip \pim$.
     The first errors are statistical, the second systematic.
        }
     \centering
  \begin{tabular}{lcc}
\hline \\ [-2.3ex]
Intermediate state & fraction (\%) & phase (rad)\cr
\hline \\ [-2.3ex]
$a_0(980)^+ \pim$ & \alm$12.3 \pm 1.2 \pm 2.8$ & 0.  \cr
$a_2(1310)^+ \pim$ & $2.5 \pm 0.7 \pm 0.9$ & $-1.04 \pm 0.13 \pm 0.20$  \cr
$f_0(500) \eta$ & $4.3 \pm 1.3 \pm 1.1$ & \al\all$0.54 \pm 0.14 \pm 0.24$ \cr
$f_2(1270) \eta$ & $4.6 \pm 0.9 \pm 0.8$ & $-1.15 \pm 0.11 \pm 0.05$  \cr
$f_0(980) \eta$ & $5.7 \pm 1.3 \pm 1.5$ & $-2.41 \pm 0.09 \pm 0.07$ \cr
$f_0(1500) \eta$ & $4.2 \pm 0.7 \pm 0.9$ &  \al\all$2.32 \pm 0.13 \pm 0.17$\cr
$a_0(1450)^+ \pim$ & \alm$15.0 \pm 2.4 \pm 3.2$ &  \al\all$2.60 \pm 0.09 \pm 0.11$  \cr
$a_0(1700)^+ \pim$ & $3.5 \pm 0.8 \pm 0.8$ &  \al\all$1.39 \pm 0.15 \pm 0.20$  \cr
$f_2(1950) \eta$    & $4.2 \pm 1.0 \pm 1.0$ & $-1.59 \pm 0.15 \pm 0.21$ \cr
\hline \\ [-2.3ex]
resonant sum &  $ 56.3\pm 3.7 \pm 10.0$  & \cr
\hline \\ [-2.3ex]
$NR$            & \alm$172.7 \pm 8.0 \pm 10.0$ &  \al\all$1.67 \pm 0.07 \pm 0.06$  \cr
\hline \\ [-2.3ex]
sum & \alm$229.0 \pm 8.8 \pm 14.1$ & \cr
\hline \\ [-2.3ex]
$p$-value & 9.3\% & \cr
\hline \\ [-2.3ex]
  \end{tabular}
\label{tab:table6}
\end{table}
A new $a_0(1700)$ resonance is observed in the $\eta \pipm$ invariant-mass spectrum, with fitted parameters listed in Table~\ref{tab:table4}. 
The likelihood change obtained when the resonance is excluded from the fit is $\Delta(-2\log \calL)=72.3$, corresponding to a significance greater than $8\sigma$.
We note the presence of a very large non-resonant scalar contribution, and in Table~\ref{tab:table6}, we list both the sum of resonant contributions and the sum including the non-resonant contribution. 
This effect could be correlated with the interference of the \etac with the two-photon continuum.

\subsubsection{Results from the $\etac \to \eta \Kp \Km$ analysis.}

\begin{figure}
  \begin{center}
    \includegraphics[width=7cm]{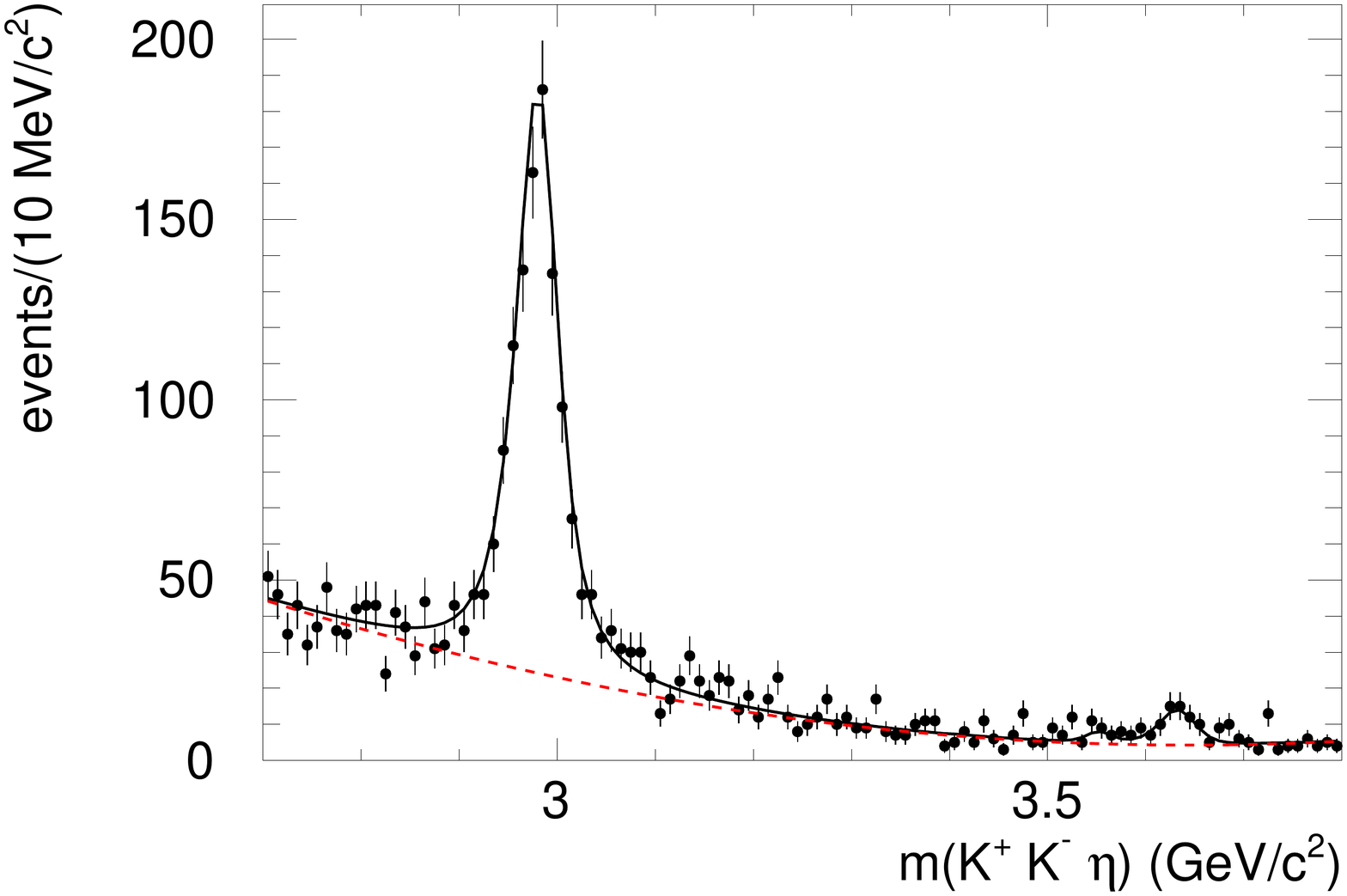}
    \includegraphics[width=7cm]{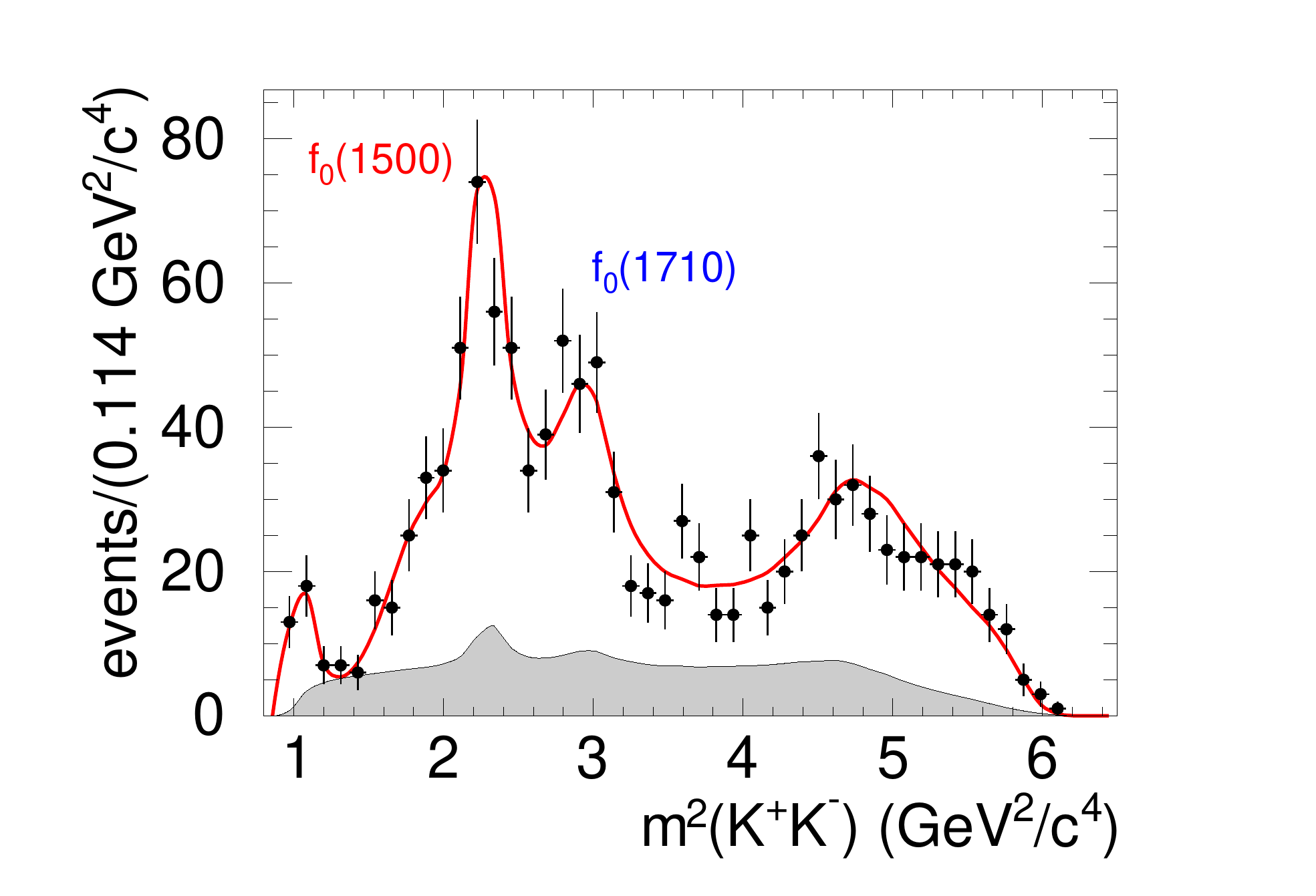}
  \caption{(Left) Invariant $\eta \Kp \Km$ mass spectrum from $\gamma \gamma \to \eta \Kp \Km$. (Right) Squared $\Kp \Km$ mass projection from the \etac Dalitz plot. The line is the result from the Dalitz plot analysis.}
\label{fig:fig11}
\end{center}
\end{figure}

To complete the list of the results summarized in the present review, we also include in fig.~\ref{fig:fig11}(Left), the $\etac \to \eta \Kp \Km$ mass spectrum combined for the $\eta \to \gamma \gamma$ and $\eta \to \pip \pim \piz$ decay modes, first observed by \babar~\cite{BaBar:2014asx}.

Figure.~\ref{fig:fig11}(Right) shows the squared $\Kp \Km$ mass projection from the \etac Dalitz plot, where signals of $f_0(1500)$ and $f_0(1710)$ can be seen.
The Dalitz plot analysis allow to measure the fractions relative to these resonant contributions which are listed in Table~\ref{tab:table7}.

\section{Conclusions}

The study of radiative \OneS decay to $\gamma \pip \pim$ and $\gamma \Kp \Km$
shows the presence of the gluonium candidates $f_0(1500)$ and $f_0(1710)$, in agreement with what observed in $J\psi$ radiative decays. 

In the framework of the identification of scalar gluonium states, it is interesting to compare the rates of \etac decays into a gluonium candidate state and an $\eta$ or an \etapr meson.

Table~\ref{tab:table7} summarizes relevant results from the analyses reported in the present review.
\begin{table} 
    \caption{\small Fractional contributions to $\etac\to \eta h^+h^-$ and $\etac\to \etapr h^+h^-$ decays of selected scalar mesons, uncorrected for unseen decay modes.
        }
    \centering
  \begin{tabular}{lccc}
\hline \\ [-2.3ex]
Final state & $f_0(1500)$(\%) & $f_0(1710)$(\%) & $f_0(2100)$(\%) \cr
\hline \\ [-2.3ex]
$\eta \Kp \Km$ & $23.7 \pm 7.0 \pm 1.8$ & \all$8.9 \pm 0.2 \pm 0.4$ & \cr
$\eta \pip \pim$ & \al$4.2 \pm 0.7 \pm 0.9$ &  & 0. \cr
$\etapr \Kp \Km$ & \al$0.8 \pm 1.0 \pm 0.3$ & \al$29.5 \pm 4.7 \pm 1.6$ & \cr
$\etapr \pip \pim$ & \alm\alm\alm\alm$0.3 \pm 0.2$ & & $74.9 \pm 7.5 \pm 3.5$ \cr
\hline \\ [-2.3ex]
  \end{tabular}
\label{tab:table7}
\end{table}
We observe an enhanced contribution of $f_0(1710)$ in \etac decays to \etapr and an enhanced contribution of $f_0(1500)$ in \etac decays to $\eta$. This effect may point to an enhanced gluonium content in the $f_0(1710)$ meson.
A similar effect is observed for the $f_0(2100)$ resonance.
The observation of $f_0(2100)$ in both $J/\psi$ radiative decays and in $\etac \to \etapr \pip\pim$ allows to add this state in the list of the candidates for the scalar glueball.


\begin{thebibliography}{6}
%
\bibitem{Chen:2005mg}
Y.~Chen, \textit{et al.}
Phys. Rev. D \textbf{73}, 014516 (2006)
[arXiv:hep-lat/0510074].
\bibitem{Minkowski:1998mf}
P.~Minkowski and W.~Ochs,
Eur. Phys. J. C \textbf{9} (1999), 283-312
[arXiv:hep-ph/9811518].
\bibitem{Amsler:1995tu}
C.~Amsler and F.~E.~Close,
Phys. Lett. B \textbf{353} (1995), 385-390
[arXiv:hep-ph/9505219].
\bibitem{Amsler:1995td}
C.~Amsler and F.~E.~Close,
Phys. Rev. D \textbf{53} (1996), 295-311
[arXiv:hep-ph/9507326].
\bibitem{Janowski:2014ppa}
S.~Janowski, F.~Giacosa and D.~H.~Rischke,
Phys. Rev. D \textbf{90} (2014) no.11, 114005
[arXiv:1408.4921 [hep-ph]].
\bibitem{Gui:2012gx}
L.~C.~Gui \textit{et al.} [CLQCD Collaboration],
Phys. Rev. Lett. \textbf{110} (2013) no.2, 021601
[arXiv:1206.0125 [hep-lat]].
\bibitem{Ablikim:2013hq}
M.~Ablikim \textit{et al.} [BESIII Collaboration],
Phys. Rev. D \textbf{87} (2013) no.9, 092009
[erratum: Phys. Rev. D \textbf{87} (2013) no.11, 119901]
[arXiv:1301.0053 [hep-ex]].
\bibitem{Kopke:1988cs}
L.~Kopke and N.~Wermes,
Phys. Rept. \textbf{174} (1989), 67.
\bibitem{Dobbs:2015dwa}
S.~Dobbs, A.~Tomaradze, T.~Xiao and K.~K.~Seth,
Phys. Rev. D \textbf{91} (2015) no.5, 052006
[arXiv:1502.01686 [hep-ex]].
\bibitem{BaBar:2018uqa}
J.~P.~Lees \textit{et al.} [BaBar Collaboration],
Phys. Rev. D \textbf{97} (2018) no.11, 112006
[arXiv:1804.04044 [hep-ex]].
\bibitem{CLEO:2005koa}
S.~B.~Athar \textit{et al.} [CLEO Collaboration],
Phys. Rev. D \textbf{73} (2006), 032001
[arXiv:hep-ex/0510015].
\bibitem{BaBar:2021fkz}
J.~P.~Lees \textit{et al.} [BaBar Collaboration],
Phys. Rev. D \textbf{104} (2021) no.7, 072002
[arXiv:2106.05157 [hep-ex]].
\bibitem{Yang:1950rg}
C.~N.~Yang,
Phys. Rev. \textbf{77} (1950), 242-245.
\bibitem{Harland-Lang:2013ncy}
L.~A.~Harland-Lang, V.~A.~Khoze, M.~G.~Ryskin and W.~J.~Stirling,
Eur. Phys. J. C \textbf{73} (2013), 2429
[arXiv:1302.2004 [hep-ph]].
\bibitem{Bass:2018xmz}
S.~D.~Bass and P.~Moskal,
Rev. Mod. Phys. \textbf{91} (2019) no.1, 015003
[arXiv:1810.12290 [hep-ph]].
\bibitem{Flatte:1976xu}
S.~M.~Flatte,
Phys. Lett. B \textbf{63} (1976), 224-227.
  \bibitem{WA76:1991kef}
T.~A.~Armstrong \textit{et al.} [WA76 Collaboration],
Z. Phys. C \textbf{51} (1991), 351-364.
\bibitem{PDG2016} C. Patrignani \etal\ (Particle Data Group), Chin. Phys. C {\bf 40}, 100001 (2016).
\bibitem{Mark-III:1986mfz}
J.~Becker \textit{et al.} [Mark-III Collaboration],
Phys. Rev. D \textbf{35} (1987), 2077.
\bibitem{BaBar:2012iuj}
J.~P.~Lees \textit{et al.} [BaBar Collaboration],
Phys. Rev. D \textbf{85} (2012), 112010
[arXiv:1201.5897 [hep-ex]].
\bibitem{BaBar:2014omp}
A.~J.~Bevan \textit{et al.} [BaBar and Belle Collaborations],
Eur. Phys. J. C \textbf{74} (2014), 3026
[arXiv:1406.6311 [hep-ex]].
\bibitem{PDG2020} P.A. Zyla \etal\  (Particle Data Group), Prog. Theor. Exp. Phys. 2020, 083C01 (2020).
\bibitem{Belle:2012uhr}
C.~C.~Zhang \textit{et al.} [Belle Collaboration],
Phys. Rev. D \textbf{86} (2012), 052002
[arXiv:1206.5087 [hep-ex]].  
\bibitem{Asner:2003gh}
D.~Asner,
[arXiv:hep-ex/0410014].
\bibitem{Blatt:1952ije}
J.~M.~Blatt and V.~F.~Weisskopf,
Springer, 1952,
ISBN 978-0-471-08019-0.
\bibitem{BaBar:2014asx}
J.~P.~Lees \textit{et al.} [BaBar Collaboration],
Phys. Rev. D \textbf{89} (2014) no.11, 112004
[arXiv:1403.7051 [hep-ex]].
\bibitem{BESIII:2014dlb}
M.~Ablikim \textit{et al.} [BESIII Collaboration],
Phys. Rev. D \textbf{89} (2014) no.7, 074030
[arXiv:1402.2023 [hep-ex]].
\bibitem{BaBar:2015kii}
J.~P.~Lees \textit{et al.} [BaBar Collaboration],
Phys. Rev. D \textbf{93} (2016), 012005
[arXiv:1511.02310 [hep-ex]].
\bibitem{Aston:1987ir}
D.~Aston, \textit{et al.} [LASS Collaboration]
Nucl. Phys. B \textbf{296} (1988), 493-526.
\bibitem{wilks} S. S. Wilks, Ann. Math. Stat. 9 (1938) 60.
\end{thebibliography}
\end{document}